\documentclass[]{aastex7}
\usepackage[utf8]{inputenc}
\usepackage{amsmath}
\usepackage{graphicx}

\shorttitle{Narrow Emission Lines of Sy1 Galaxies}
\received{July 30, 2025}
\accepted{November 21, 2025}
\submitjournal{ApJ}

\begin{document}

\title{The Narrow Emission Lines of Seyfert 1 Galaxies: Comparisons with a Large SDSS Sample}

\author[0000-0001-6919-1237]{Malkan, Matthew A.}
\affiliation{Department of Physics and Astronomy, University of California, 475 Portola Plaza, Los Angeles, CA 90095, USA}
\email[show]{malkan@astro.ucla.edu}

\author{Jensen, Lisbeth D.}
\affiliation{Department of Physics and Astronomy, University of California, 475 Portola Plaza, Los Angeles, CA 90095, USA}
\email[show]{lizjensenmusic@gmail.com}

\author[0000-0003-2478-9723]{Hao, Lei}
\affiliation{Shanghai Astronomical Observatory, Chinese Academy of Sciences, 80 Nandan Road, Shanghai 200030, People’s Republic of China}
\email[show]{haol@shao.ac.cn}



\begin{abstract}

We analyzed a large sample of SDSS spectra of Seyfert galaxies, sub-dividing Seyfert1’s based on their narrow-to-broad H$\alpha$-components. Comparing their narrow-lines (NL) to Seyfert2s in line-ratio diagrams, most of the NL of strong Sy1.0/Sy1.2's (with dominant broad lines) are the same as those of ``pure" Sy2s. In contrast, only $\sim$25-30\% of the Sy1.8 and Sy1.9 nuclei (with weak broad lines) are located in the pure Sy2 region, with the rest falling in the composite/star-forming region. We explain these Seyfert-plus-star-formation spectra with a simple model. It shows that 85\% of NL in Sy1.9 are from HII-regions, while 88\% of the NL in Sy 1.0 arise from the same NLR as in pure Sy2. About $\sim$6\% of the strong and weak Sy1's have NL dominated by LINER emission, while $\sim$15\% of intermediate Seyferts (Sy1.5/Sy1.6) do.

To confirm this Seyfert 1 AGN-plus-star-formation combination, we used stellar absorption-lines to compare their stellar populations. Their H$\delta$ strengths show that LINERs, pure Sy2’s, and also the broad-line-dominated Sy1’s have old stellar populations. The weak Sy1’s show stronger H$\delta$ 
absorption, indicating larger proportions of young stars. About one-third of the u-band light in Sy1.0/1.2 is blended Balmer-lines and continuum from the BLR. The NL gas reddening increases as the BLR strength decreases, from Sy1.0(0.13mag), to Sy1.9(0.40mag), to Sy2s and LINERs both with 0.50mag. Our data do not support the simplest version of Seyfert 1 and 2 unification, where both AGN classes have identical NL.
\end{abstract}



\section{Introduction} 
\label{Introduction1.0}

The emission line spectra of Type 1 Seyfert active galactic nuclei (AGN) are defined by having at least some broad permitted lines. Even when they are relatively weak, at a minimum the broad wings ($\Delta V>$ 1000 km/sec) of H$\alpha$ are detected. That is the defining characteristic of ``Seyfert Type 1" AGN classification (\cite{1992ApJS...81...59R, 1997ApJS..112..315H}). The high observed velocities of the ``broad-line" gas cannot be produced in normal galaxies, and are instead caused by large accelerations (due to gravity and possibly radiation pressure, see \cite{1999ApJ...526..579W}) produced uniquely near ($R_{blr} < 1$ parsec) the massive central engine (\cite{1992ApJ...392..470P,2013ApJ...767..149B}), defined as the ``broad-line region" (BLR).

A separate, somewhat larger class of AGN, the Seyfert 2 galaxies, are defined by having {\it no directly observable broad permitted emission line wings} \citep{1983ApJ...265...92M}. They are instead identified by their strong ``narrow" emission lines ($FWHM < 1000$ km/sec). Although their lines arise at larger radii in a so-called ``Narrow Line Region" (NLR; $R_{nlr} \ge$ 10 pc), they nonetheless show {\it highly ionized gas} that is photoionized by the hard UV/Xray continuum of the AGN \citep{2022ApJ...941...46S}. The softer ionizing spectra of O stars do not produce such high ionization states as are seen in the NLR.  These narrow AGN lines are forbidden transitions, which are not emitted in the BLR due to collisional suppression at high densities ($n_e \ge 10^8 cm^{-3}$; \cite{1992ApJ...399..504S}). 

But it has long been recognized that Seyfert 1 spectra also usually include strong narrow emission lines,
which in many cases arise from the same kind of highly
ionized gas that produces the same narrow lines in
Seyfert 2 nuclei.  Indeed, the compelling hypothesis that Seyfert 1 = Seyfert 2---Plus-Broad-line region
was the first effort at observational ``unification" of the active galactic nuclei (c.f., \cite{1983ApJ...265...92M}).

A major observational challenge is
that the permitted emission lines, such as the hydrogen recombination lines, simultaneously show
{\it both broad and narrow components superposed} in Seyfert 1 spectra. Separating the broad and narrow
components of the two strongest H lines--H$\alpha$ and H$\beta$--requires careful
analysis of spectra with both high spectral resolution and high SNR
(\cite{1990ApJ...350..132R, 1992ApJ...392..470P, 2005AJ....129.1783H}).
Our previous analysis of AGN emission lines (\cite{2017ApJ...846..102M}; M17 hereafter) did not in general have such spectroscopy and therefore concentrated mostly on the forbidden line and the total Balmer line fluxes (broad-plus-narrow components.)
This limitation has been overcome through the
availability of large samples of high-quality galaxy spectra--by far the largest being produced by the Sloan Digital Sky Survey (SDSS), Seyfert 1 galaxies from \cite{2005AJ....129.1783H} (hereafter H2005).
In particular the SDSS spectra allow a reliable separation of the narrow and broad components
of H$\alpha$ and H$\beta$. 

The starting point for this paper is the dataset of SDSS spectra of AGN
assembled and measured by one of us (H2005). 
Using the same procedures she had applied to the Data Release 2 of the SDSS in 2005, this new sample has included many more AGN, since it is based on all the emission line galaxy spectra available in Data Release 7. See \S \ref{section2.0} for an in-depth description of our dataset. 

\subsection{The nature of the narrow line emission from Seyfert 1 galaxies}\label{Introduction1.1}

For several decades it has been supposed that Seyfert 1's differ only from Seyfert 2's in their broad permitted emission lines. 
The simple version of ``unification" 
assumes that there is {\it no difference} in their NLRs. Operationally, the NLR emission lines have observed velocity widths FWHM $<$ 1000 km/sec  (and usually less than 700 km/s).  This same gas should emit Hydrogen Balmer emission lines with similar ``narrow" velocity width. And indeed in many Seyfert 1 spectra, the central cores of the Balmer emission lines do show a distinctive ``narrow" component that stands out above the apparently separate ``broad" permitted line emission  ($FWHM >$ 1000 km/sec; e.g. \citep{1992ApJS...81...59R, 1994ApJS...93...73R}). 
This is a robust separation, since the velocities of the NLR and BLR are so different, and the narrow lines never show time variability, unlike the broad lines which frequently do. This is because the narrow lines arise in a far larger NLR volume \citep{1993ApJ...408..416D,1998ApJ...509..163O,1992ApJ...392..470P}.

Nonetheless, there were already reasons to doubt that all NLR's are identical in all types of Seyfert galaxies. H2005, M17, and others pointed out that this simple 2-component analysis is inadequate to describe some details of the emission-line spectra. In particular, the SDSS spectra, obtained through 3-arcsec diameter fibers,
can include additional narrow line components, either from HII regions
in the disk of the host galaxy, or from extended relatively low-ionization gas
which is often seen in galaxies, especially those of early-type, 
producing a LINER spectrum {\it Low-Ionization Nuclear Emission Line Region }\citep{1980A&A....87..152H}.
Thus in Seyfert galaxies of both Type 1 and 2, the narrow emission
lines may in general arise from at least 3 physically distinct components--the AGN-powered
NLR, HII regions powered by young hot stars, and a possible LINER. 

In this paper, we test the importance of additional narrow-line components in the spectra of Seyfert 1 galaxies.  
To put this another way, by subtracting away the broad permitted lines to analyze the remaining narrow-line spectrum we are asking: 
What would the spectrum of this Seyfert 1 galaxy look like {\it after} its broad-line region had completely shut down for a few centuries?

This paper is organized as follows. In the following section (\S \ref{section2.0}) we describe our data and how it was processed. In section \S \ref{section3.0} we describe how the data sample was divided into several sub-groups, and investigate how these groups fall in the traditional line-ratio diagrams as well as in an alternate diagram introduced by \cite{2006MNRAS.372..961K}. Section \S \ref{section4.0} discusses the prediction of [NII] from the [SII] doublet. In sections \S  \ref{section5.0}, \ref{section6.0}, and \ref{section7.0} gas reddening, electron densities, and ionization parameters are estimated and compared, respectively. Sections \S \ref{section8.0} and \ref{section9.0} we analyze and interpret absorption line measurements of the starlight continuum in our categories of AGN spectra, and possible contamination from AGN components. Finally, in section \S \ref{section10.0} we discuss the implications of our results for selection of Seyfert galaxy samples, and the standard unification theory for Type I and Type II Seyferts. Our conclusions are summarized in \S \ref{section11.0}.

\section{The AGN and Star-Forming Galaxy Sample}\label{section2.0}
Our AGN emission-line spectra sample was obtained from H2005, hereafter referred to the ``Hao-sample") who initially extracted Galaxy-Target objects and Quasar-Targets with $r(petrosian) \leq  17.99$ from the SDSS DR2 catalog, and later extended this sample to include AGNs from SDSS DR7 (previously unpublished data). We supplement these data with additional emission and absorption line flux measurements, such as the [NeIII]3869 and D4000 lines, from the well-known MPA-JHU DR7 catalog.
\footnote{From \url{https://wwwmpa.mpa-garching.mpg.de/SDSS/DR7/}}
\footnote{\cite{2012MNRAS.419.1402G} estimates that H$\beta$ EW is underestimated by $<\sim0.35>$\AA, 
in the MPA-JHU DR7 sample, corresponding to $\sim$ 5\% systematic vertical shift in the BPT diagrams. Since we do not use H$\beta$ from this catalog, except for location of the SF galaxies, we have not applied this correction.}
\citep{2009ApJS..182..543A}, hereafter referred to as the ``MPA-catalog". Both the Hao-sample and the MPA-catalog have the Plate-ID, Fiber-ID, and MJD listed, and we use these identifiers to match Hao's AGN with the galaxies in the MPA-catalog with the \text{TOPCAT} astronomy software \citep{2005ASPC..347...29T}. 

From the MPA-catalog we also select a large sample of star-forming (SF) galaxies used in our analysis.
We require that all galaxies have a redshift $\leq$ 0.33 to ensure good measurements of the key red emission lines within the SDSS optical spectral range, and we exclude all galaxies with emission line fluxes having S/N $<$ 5. Since H$\beta$ is usually the weakest, our sample is--in effect-- set by S/N(H$\beta$) $\ge$ 5.
The original sample from Hao comprised 23,157 galaxies classified as narrow-line galaxies of both Seyfert 2 and LINERs, and 10,015 broad-line (Seyfert 1) galaxies. After applying our S/N ratio limit, there are 9053 narrow-line galaxies and 4097 broad-line galaxies. We separate both the narrow- and broad-line galaxies into further sub-groups, described in subsequent subsections. 

Below we summarize the procedure H2005 used to measure broad Balmer line components to separate the Seyfert 1 and Seyfert 2 AGNs, and then within the Seyfert 1 sample further separate the NLR from the BLR.  Similar AGN identification procedures have been more recently applied to other large spectroscopic datasets, which reach galaxies fainter than those in our SDSS sample.  However, for the statistical analyses described in this paper, the numbers of these new samples are still small. For example, the GAMA survey of galaxy spectra down to r=19.8 mag has spectroscopically identified a total of 954 Seyfert 1+Seyfert 2 galaxies \citep{Gordon2017}. This limitation should be overcome in the foreseeable future.

\subsection{Hao-Sample Processing procedure}\label{section2.1}
We here provide a brief summary of the emission-line measurements. Please see sections \S4 and \S5 in \cite{2005AJ....129.1783H} for a complete description.   

Because the SDSS spectra are observed through a 3\arcsec\ aperture, both nuclear and stellar-light from the host galaxy are mixed into each spectrum. Since the starlight component is a continuous spectrum with absorption lines, H2005 built a set of absorption-line spectral templates and using principal component analysis (PCA) they removed the underlying stellar continuum, leaving only the pure emission-line spectra. H2005 then removed from their sample any spectra with a (starlight-subtracted) equivalent width of H$\alpha$ less than 3\AA. Our more stringent 5-$\sigma$ requirement for H$\beta$ detection, removes almost 60\% of the original Seyfert sample. This heavy restriction was necessary to ensure that the H$\alpha$ measurements we use are accurate.  Our median observed equivalent width of total H$\alpha$ emission is 18\AA, with a very sharp drop-off below 8\AA.  Although our final sample excludes weak emission-line galaxies, we can be confident that the H$\alpha$ components we do analyze are accurately measured. 

In each SDSS spectrum, the FWHM, intensity, central wavelength, and surrounding continuum of H$\alpha$, H$\beta$, [OIII]$\lambda$5007, the [SII]$\lambda\lambda$6716+6724 and [OII]$\lambda\lambda$3726+3729 doublets, and [OI]$\lambda$6300 emission-lines are measured with Gaussian-profile fits. Measuring the FWHM width of a single-Gaussian fit to H$\alpha$ naturally splits narrow and broad lines into a bi-modal distribution with a clear separation at 1200 km s$^{-1}$ (figure 5 in H2005). This same sharp velocity width distinction was found by \cite{Vaona2012} in their analysis of 2600 Seyfert galaxy spectra from SDSS-DR7.
The median ``narrow" line FWHM is 320 km s$^{-1}$. This procedure separated the Hao-sample into two general galaxy groups, one containing narrow-line regions (NLR) (Seyfert 2s and LINERs), and the other with broad line regions (BLR) (Seyfert 1s).

\subsection{Separation of the Narrow Line Region in the Broad Line Seyfert 1 Galaxies}\label{section2.2}
The permitted broad-lines in Seyfert 1 spectra produced by high-velocity gas are often seen to be superposed on low-velocity narrow cores. As mentioned in \S1, the large line-widths in the Sy 1 BLR are caused by large accelerations within a few parsecs of the central engine, while the NLR originates much further out. Both regions are generally contained within the SDSS fiber spectra. Thus, the combined high- and low-velocity components of permitted lines  such as H$\alpha$--from the BLR and NLR respectively--can only be separated by careful spectral fitting. Since this paper focuses on this separation, we summarize this procedure below.

To identify galaxies containing both broad and narrow permitted line components, the narrow H$\alpha$ plus the [NII]($\lambda\lambda$6584+6548) doublet is first fitted with a three-Gaussian model. The key test for discovering Seyfert 1 nuclei is to add a fourth Gaussian to model both the narrow and broad H$\alpha$ components. If this gives a substantially better fit than any 3-component Gaussian, it is determined to be a Sy 1 galaxy. The quantitative criterion for preferring the four-Gaussian model for any given galaxy comes from comparing the  $\chi^{2}$ of each model:
\begin{equation}\label{eq:1}
    (\chi^{2}_{3} -  \chi^{2}_{4} - 2)/\chi^{2}_{4} > 0.2
\end{equation}
where subscripts 3 and 4 indicate the respective models. The addition of 2 more degrees of freedom for the four-Gaussian model is expected to reduce the $\chi^{2}$. The numerical limit of 0.2 is explained by H05 as ``an empirical number and is demonstrated to be appropriate by manual inspection". This criterion is built on statistical fitting models by \cite{Lupton1993}.

One limitation of this method is that the narrow lines often have non-Gaussian wings. For a strong narrow-line with these extended wings, the four-Gaussian model is used (given by equation \ref{eq:1}). However, H2005 prefers the three-Gaussian fit if the height of the narrow-line component ($h_n$) is large compared to the broad-line height $h_b$ and the width of this line $\sigma_w$ is small:
\begin{equation}\label{eq:2}
    \sigma_b < 20\text{\AA} ,\; \frac{h_b}{h_n} < 0.1
\end{equation}

The reason why the narrow and broad components of H$\alpha$ can be separated cleanly in most Seyfert 1 spectra is that the FWHM values of most narrow and broad components are so extremely different, with virtually no overlap.

\begin{deluxetable}{lcccccc}	
\tablenum{1}					
\tablecaption{Classification of Galaxies Spectra}		
\tablewidth{0pt}					
\tablehead					
{					
\colhead{Galaxy} & 
\colhead{Flux Ratio ($R_{H\alpha}$)} & 
\multicolumn{2}{c}{H${\alpha}$ FWHM Median} &
\colhead{Avegage}	& 
\colhead{Hao} & 
\colhead{MPA} \\
\colhead{Group} &
\colhead{(H$\alpha$ N)/(H$\alpha$ B)} &
\colhead{Narrow Line} &
\colhead{Broad Line} &
\colhead{Redshift} & 
\colhead{Sample$^a$} & 
\colhead{Sample$^b$}
}					
\startdata					
Sy 1.0 	&	 $\qquad \quad R_{H\alpha} \le 0.25$ 	&	339	&	4420	&	 0.102	&	677	&	135\\
Sy 1.2 	&	 $0.25 < R_{H\alpha} \le 0.50$ 	&	336	&	3390	&	 0.104 	&	846	&	315\\
Sy 1.5 	&	 $0.50 < R_{H\alpha} \le 1.00$ 	&	311	&	2430	&	 0.110	&	1187	&	713\\
Sy 1.6 	&	 $1.00 < R_{H\alpha} \le 1.50$ 	&	308	&	2100	&	 0.105	&	662	&	500\\
Sy 1.8 	&	 $1.50 < R_{H\alpha} \le 2.50$ 	&	319	&	2320	&	 0.113  	&	420	&	336	\\
Sy 1.9 	&	 $\qquad \quad R_{H\alpha} > 2.50$ 	&	264	&	2530	&	 0.096&	208	&	164\\
Sy 2.0 Lower 	&	 ... 	&	 ... 	&	 ... 	&	  0.099	&	3045	&	1573\\
Sy 2.0 Upper 	&	 ... 	&	 ... 	&	 ... 	&	  0.109	&	4783	&	3510\\
LINERs 	&	 ... 	&	 ... 	&	 ... 	&	0.100 	&	1275	&	976	\\
SF-Lower 	&	 ... 	&	 ... 	&	 ... 	&	 0.096  	&	 ... 	&	70801\\
SF-Middle 	&	 ... 	&	 ... 	&	 ... 	&	 0.089 	&	 ... 	&	62479\\
SF-Upper 	&	 ... 	&	 ... 	&	 ... 	&	  0.068	&	 ... 	&	16675\\
\enddata
\tablecomments{The Seyfert 1 sub-groups are defined according to the Flux ratio of H$\alpha$ NLR to the H$\alpha$ BLR, while ensuring that the FWHM of the Narrow Line component is less than 1200 Km s$^{-1}$. The Sy 2.0 Lower and Upper, LINERs, and starforming sub-groups are defined according to the BPT diagram shown in Figure \ref{fig:fig2}. The redshift of all Seyferts and LINERs is $<z> \sim 0.1$; thus different group redshifts cannot be a source of any bias. SF-Upper has a lower $<z>$ compared to SF-Middle and SF-Lower. \\
$^a$ Number of Seyfert and Liner Galaxies from Hao sample. \\
$^b$ Number of Seyfert and Liner Galaxies from the Hao sample, which are also in the MPA-JHU catalog.}\label{table:1}	
\end{deluxetable}

\subsection{Seyfert 1 Sub-Classes from Relative Strengths of Broad and Narrow Balmer Lines}\label{section2.3}
A key parameter in classification of Seyfert 1 spectra is the ratio of the BLR to NLR contributions of permitted emission lines. This sub-classification was pioneered by observers at Lick Observatory, who first defined "Seyfert 1.5's", then "Seyfert 1.9's and 1.8's" as having a relatively stronger NLR contribution, and then a heavily dominant NLR contribution. In particular they \citep{1981ApJ...249..462O} defined Sy1.8's to have only very weak broad wings under H$\alpha$ and H$\beta$, while Sy1.9's have only broad wings under H$\alpha$,  but no perceptible broad excess H$\beta$. To capture these distinctions, we have devised a quantitative spectral classification scheme using only the H$\alpha$ narrow-to-broad flux ratio, defined as: 

\begin{equation}\label{eq:3}
R_{H\alpha} = \frac{H \alpha (Narrow)}{H \alpha (Broad)}
\end{equation}
The Sy 1 sub-categories are based on ranges of this ratio, for example we define the Sy 1.0 as having $R_{H\alpha} < 0.25$ and for the Sy 1.9 $R_{H\alpha} > 2.50$. 
In this paper we also identify an additional sub-group with weaker broad H$\alpha$ than Seyfert 1.5s, which we label ``Seyfert 1.6". Admittedly, our adjacent Sy 1.n categories share similarities, but there are enough significant differences to justify our separating out 6 distinct categories. All Seyfert sub-categories and definitions are given in Table \ref{table:1}, with the total number of galaxies in each group.  For comparison, this table includes the total number of Sy 2 Lower/Upper, LINERs, and number of Star-Forming galaxies, along with their average redshifts.
\footnote{We note that our Seyfert 1 subcategories are based on the {\it relative} dominance of the BLR. They are {\it not an absolute luminosity sequence}, since all of them have the same average redshift and average [OIII] luminosity--$4 \times 10^7 L_{Sun}$. }

This simple adopted definition based on $R_{H\alpha}$ for all of our Seyfert 1 galaxies deliberately ignores information about the profile of the H$\beta$ emission line. We found this limitation necessary because most of the SDSS spectra we are analyzing in this paper lack sufficient signal-to-noise ratios to provide reliable information about detailed profiles of H$\beta$.
Fortunately, the simple quantitative separation of the narrow and broad components of the much stronger H$\alpha$ line emission now allows us to make this classification for thousands of Seyfert 1 galaxies--the main basis of this work shown in Table \ref{table:1}
\footnote{The strong emission lines used in our various classification diagrams have typical individual uncertainties smaller than 0.1 dex.  However, for a given Sy1 sub-class, the observations show considerably larger scatter. One possible explanation for some of this scatter is that the flux of the broad H$\alpha$ component is variable on timescales of years, while the much more extended forbidden emission line fluxes remain constant. This variability could move an individual Seyfert 1 galaxy from one Sy1.n sub-class to another without any other observable changes \citep{1993ApJ...408..416D, 2022ApJ...925...52U, 2007ApJ...661...60W, 2016ApJ...821...33R}.}.

\begin{figure*}[ht!] 
\includegraphics[width=1.0\textwidth]{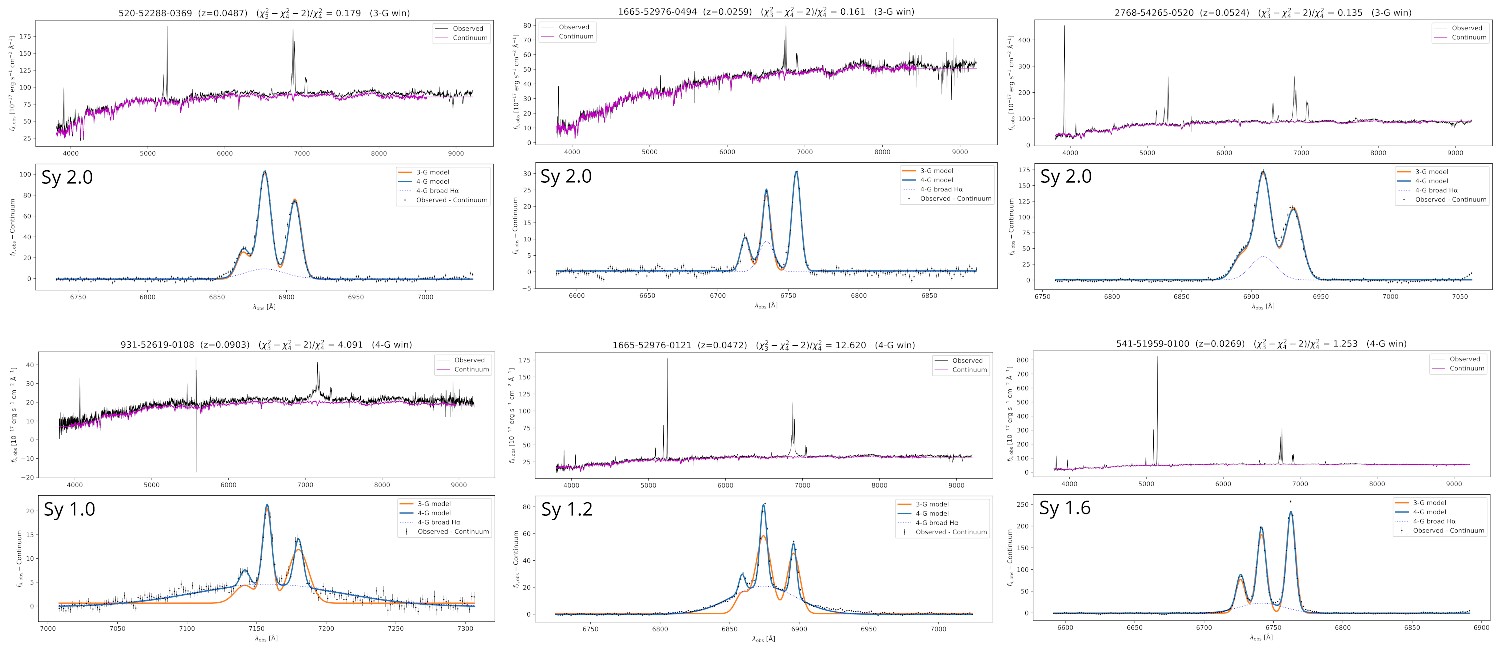}
\caption{ Top Two Rows show three examples of Seyfert 2 galaxies with the full spectra in blue and the fitted continuum in red. In the second row of spectra zoomed in on the H$\alpha$/[NII] region, show that the reduction in $\chi^2$ when adding a fourth Gaussian to the emission line fit is not statistically significant.  
The bottom two rows show three examples of Seyfert 1 galaxies, with the 3-Gaussian fit shown in red and the (preferred) 4-Gaussian fit in blue. The reduction in $\chi^2$ is large enough to require the inclusion of broad components in their H$\alpha$ lines.  Based on our quantitative comparison of narrow-to-broad-line H$\alpha$ components, these are classified as Seyfert 1.0, 1.2, and 1.6.}
\label{fig:fig1}
\end{figure*}

The top rows in Figure \ref{fig:fig1} show examples of three Seyfert 2 galaxies for which the addition of a fourth Gaussian, for broad H$\alpha$, does not significantly improve the fit, and where equation \ref{eq:1} is less than 0.2. The bottom rows show examples of three Seyfert 1 galaxies in which a broad H$\alpha$ component is statistically required for a good fit. These galaxies have a range of narrow/broad H$\alpha$ fluxes corresponding to Sy 1.0, 1.2 and 1.6 sub-classifications, and equation \ref{eq:1} is 4.09, 12.62, and 1.25, respectively. 

To check our simple Seyfert 1 sub-classification, we have visually examined a representative subset of these SDSS Seyfert 1 spectra from our sample. We confirm that in most cases the decimal integer sub-class after the ``1" does agree {\it qualitatively} with what the original Lick observers would have assigned. However, we caution that compared to many of their ``classical" Seyfert 1 nuclei, which dominate the literature over the previous several decades, the Hao-sample Sefyert 1 spectra tend to have considerably weaker broad H Balmer lines. Indeed, most of our SDSS AGN which our equation \ref{eq:3} classifies as Sy 1.6 or Sy 1.5, and even some of the Sy 1.2s, do not show {\it clear} broad H$\beta$ components in their spectra.  They would therefore all be classified as Seyfert 1.9's in the original Lick system. Thus we do not expect that most of the SDSS Seyfert 1's in our study will be {\it quantitatively} identical to the first original samples of Seyfert 1's studied in earlier publication, which tended to be more dominated by their broad H emission lines. 
In studying trends, we will sometimes refer to the broad-line dominated Seyfert 1's (1.0's and 1.2's) as ``strong Seyfert 1s", and the narrow-line dominated Seyfert 1's (1.8's and 1.9's) as ``weak Seyfert 1s". Much of the original literature on Seyfert 1 spectra refers to strong Seyfert 1's.

\subsection{Narrow-Line AGN Sample: Seyfert 2 and LINERs}\label{section2.4}
To compare the NLR of the Sy 1 sub-types with those of Seyfert 2s and LINERs, we use the time-tested ``BPT" double-line ratio diagrams \cite{1981PASP...93....5B}. In addition to the traditional  diagrams that use either [NII]/H$\alpha$, [SII]/H$\alpha$, or [OI]/H$\alpha$ on the x-axis, and [OIII]/H$\beta$ on the y-axis, we also use an alternate ``BPT diagram" introduced by \cite{2006MNRAS.372..961K} that replaces the H$\beta$ with the [OII] doublet (labeled {\it Ke06} in this paper). These optical line-ratios efficiently separate the narrow-line contribution from three types of gas ionization: Hot Stars, AGN, and LINERs. These diagrams contain a boundary line for the theoretical upper limit of starbursts \citep{2001ApJ...556..121K} (hereafter Ke01)--galaxies lying above and to the right of the Ke01 boundary are considered to be AGN- dominated (Figures \ref{fig:fig2} - \ref{fig:fig5}). \cite{2003MNRAS.346.1055K} implemented an additional boundary-line in the [NII] BPT diagram (hereafter Kaf03) that empirically separates pure star-forming galaxies from those with a combined AGN-plus-HII region spectrum, referred to as ``composite galaxies". All galaxies to the left of the Kaf01-line are star-forming, and these are the galaxies in our MPA-sample, discussed further in \S 2.5 (Figure \ref{fig:fig2}). 
\footnote{A limitation of the SDSS optical spectra is that they do not generally cover the
strong emission lines produced by more highly ionized gas than O++. In particular the stronger and more highly ionized NLR emission lines observed in the mid-infrared (e.g. O+++ and Ne++++) are purely powered by AGN photoionization \citep{2023ApJ...942L..37A}. They are ideal, unambiguous measures of the AGN component, and in fact have luminosities which are linearly correlated with the continuum luminosity of the AGN, independent of star formation in the host galaxy \citep{2015ApJ...799...21S, 2016ApJS..226...19F,2016MNRAS.458.4297G, 2008ApJ...676..836T,2010ApJ...709.1257T}.}

The narrow-line galaxies in the Hao-sample include both pure Sy 2 and LINER galaxies. We follow M17 in separating these two types using a boundary line that represents a 50\% LINER/Sy2 demarcation in the [NII] BPT diagram. It is important to note that AGNs show a continuous mix of NLR and LINER components, and also of broad-line components--there is no clear-cut separation between any of these components. In M17, to determine the relative contribution from each galaxy spectral type Sy1, Sy2, or LINERs, we computed the distance ($d_i$)  of each galaxy in this diagram from the three values representing the ``pure" Seyfert 1, 2, and LINER types. The contribution of each component, $C_i$, is defined as
\begin{equation}\label{eq:4}
C i = \frac{1/d_i}{\sum (1/d_i)}    
\end{equation}
where $i$ refers to Seyfert 1, 2, and LINER emission
components, and
\begin{equation}\label{eq:5}
\begin{split}
    d^2_i & =\left[\log\frac{[OIII]}{H\beta}-\left(\log\frac{[OIII]}{H\beta}\right)_i \right]^2 \\
    & + \left[\log\frac{[NII]}{H\alpha}-\left(\log\frac{[NII]}{H\alpha}\right)_i \right]^2
\end{split}
\end{equation}
where the average values are ($\log{([OIII]/H\beta)}$)= (-0.1, 0.85, 0.1) and ($\log{([NII]/H\alpha)}$)= (-1.1, 0.05, 0.2) for pure Sy1, Sy2, and LINERs, respectively.

The dotted line in Figure \ref{fig:fig2} shows the Sy2/LINER ``mixing curve”, with  50\% contributions from both LINER and Sy2 components, and zero broad-line component, is given by: 
\begin{equation}\label{eq:6}
    log\frac{[OIII]}{H\beta} = \frac{-1.2}{(log\frac{[NII]}{H\alpha} + 1.1)} + 1.5\, ,\,log\frac{[NII]}{H\alpha} \ge -0.15
\end{equation}

Unless a galaxy spectrum shows a $>$50\% contribution from the LINER component, it cannot be reliably classified as LINER-dominated. Any emission-line galaxy lying above this equation \ref{eq:6} line is considered Seyfert 2-dominated. See \S \ref{section3.3} for further LINER details.

We further separate the Seyfert 2 sample into two groups,
based on their vertical locations ([OIII]/H$\beta$) in the [NII] BPT diagram (Figure  \ref{fig:fig2}). Our motivation for separating these Sy2 groups is the wide range of emission line ratios observed in the large category of Sy2's. We use the [NII] BPT diagram to sub-divide the Seyfert 2's into an ``Upper" (Sy2U) and a ``Lower" (Sy2L) group, depending on whether their [OIII]/H$\beta$ ratios are above or below the dash-dot line indicated in Figure \ref{fig:fig2}a. This line starts at the 50\% LINER/Sy2 demarcation line, and is close to equidistant from the Ke01 AGN/SF separation line and is given by:
\begin{equation}\label{eq:7}
    log\frac{[OIII]}{H\beta} = \frac{0.98}{(log\frac{[NII]}{H\alpha} - 0.8)} + 1.65\,, \,log\frac{[NII]}{H\alpha} \leq 0.0
\end{equation}

Values of $log([OIII]/H\beta)$ significantly above the right- hand side of equation \ref{eq:7} indicate that a classic AGN-powered NLR {\it dominates} the [OIII] emission. Our suspicion was that the Sy2U sub-class are more dominated by {\it pure} AGN NLR emission, whereas the Sy2L sub-class has more low-ionization 'contaminations', which pull them down closer to the Ke01-line.  Our subsequent analysis confirms and quantifies this supposition (see sections \S \ref{section5.0},\S\ref{section6.0}, and \S \ref{section7.0}
).
\footnote{ These two Sy2 sub-classes have the same average redshifts. Although, by design the Sy2U class of course has more powerful [OIII] emission than the Sy2L's ($4 \times 10^7 L_{Sun} $  mean versus $1.4 \times 10^7$, both groups have the same mean H$\alpha$ luminosities, $2.4 \times 10^7$. } 

\subsection{Comparison Samples of Star-Forming Galaxies}\label{section2.5}
\begin{figure*}[ht]    
\includegraphics[width=\textwidth]{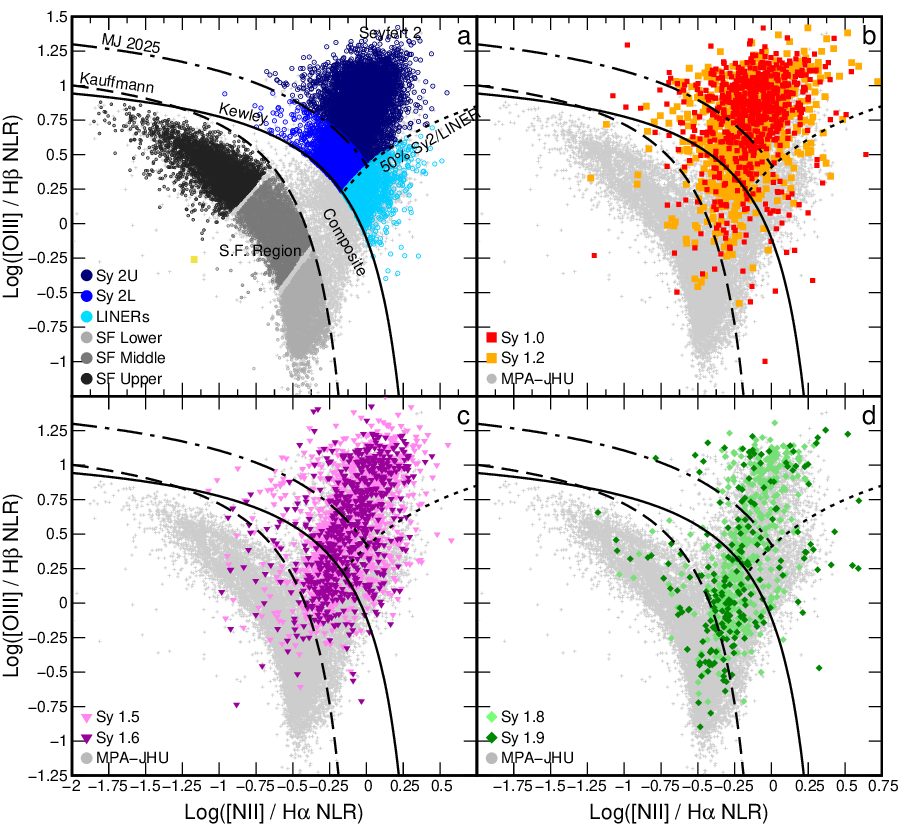}
\caption{BPT Diagrams with log([NII]/H$\alpha$) on the x-axis. In all the line ratio classification diagrams, the {\it light Grey} dots show the full MPA-JHU SDSS DR7 data. 
(a) Shaded blue data points are from the Hao AGN sample. We separate the Star-Forming region into three sections: {\it SF-Lower} bottom region--light gray, {\it SF-Middle} medium--gray, and the {Black dots} we call {\it SF-Upper}. These groups correspond to high, medium and low-metallicity galaxies, respectively.
 (b) $Red$ and $Orange$ points are Sy 1.0 and Sy 1.2 galaxies. More than 85\% of the Sy 1.0 and 77\% of the Sy 1.2 are above the Ke01 line, 
 which we consider to be dominated by the AGN NLR.
 (c) $Purple$ and $Pink$ points are Sy 1.5 and Sy 1.6 galaxies. 65\% are found in the ``Seyfert" region 
above the Ke01 boundary line
 (d) $Light\;Green$ points are the Sy 1.8 NLR, which have similar  distribution as in plot (b) but the majority of the Sy 1.9 NLR ($Dark \;Green$ points) lie in the composite region and down towards SF Lower. Only 50\% of the Sy 1.8 and 30\% of the Sy 1.9 are in the Seyfert-dominated region.}\label{fig:fig2}
\end{figure*}

The narrow and broad-line AGN sample from H2005 does not contain any galaxies with emission lines predominantly arising from SF. We therefore constructed a separate comparison SF sample from the MPA-JHU spectroscopic catalog. 
In the classical [NII] BPT diagram, the SF region can be classified as being to the left of and below the Kaf03 line, which empirically separates the SF region from composite galaxies  (Figure \ref{fig:fig2}a). These galaxies span a wide range in the [OIII]/H$\beta$ ratio, and also to some degree in [NII]/H$\alpha$. This is indicative of a wide range of ionizations and metallicities, with the lower-[OIII] galaxies having more metal-rich gas. To capture this large variation, we further sub-divide the SF region into 3 separate groups.  Figure \ref{fig:fig2}a shows how the 3 sub-groups are defined. The ``SF-Lower", ``SF-Middle", and ``SF-Upper" groups roughly correspond to high gas metallicity,  moderate metallicity, and low metallicity, respectively. In the following analysis, we separately measure each of the 3 SF sub-categories in the same spectroscopic diagnostics that we measure in our AGN sub-groups, to enable quantitative comparisons. These comparisons will show that the binary classification of "SF" versus "AGN" is overly simplified, and that many real AGN also include some (but not dominant) contributions to their emission line spectra from recent SF (\cite{2003MNRAS.346.1055K}, M17).

\section{Narrow Emission Line Ratios in Seyfert 1 Spectra}\label{section3.0}
As noted in section \S \ref{Introduction1.1}, it has been widely assumed that any narrow lines present in a Seyfert 1 spectra are produced by the same ionization mechanism and originate from the same location as in a Sy2 AGN. In all BPT diagrams as well as in the Ke06 diagram, the Sy2s are clustered toward the upper right. Thus, if the NLR Sy1s are all AGN-powered, they should position themselves at the same location. But if these NLRs are contaminated by HII-regions or LINER gas, their positions in the diagrams will be shifted downwards. 

\begin{deluxetable}{lcccccc}						
	\tablenum{2}					
	\tablecaption{Fraction (\%) of Sy 1 Galaxies falling in each region in the BPT diagrams.}
        \tablewidth{0pt}					
	\tablehead					
	{					
	\colhead{Region}
        &	\colhead{Sy 1.0}	
        &	\colhead{Sy 1.2}
        &   \colhead{Sy 1.5}
	&	\colhead{Sy 1.6}
        &	\colhead{Sy 1.8}	
        &	\colhead{Sy 1.9} 
	}					
	\startdata	
          &		&	      &	[NII] BPT &		    &	      &     \\  
\cline{2-7}
SF Region &	1.8	&	2.1	    &	3.8	    &	4.9	    &	8.9	  & 26.9\\
Composite &	8.1	&	16.8	&	22.5	&	24.0	&	34.0  & 34.8\\
Sy2L	  & 16.0&	26.1	&	23.7	&	17.3	&	13.9  & 10.2\\
Sy2U	  & 70.0&	51.2	&	41.1	&	43.9	&	36.8  & 18.9\\
LINERs	  &  4.1&	3.8	    &	8.9     &	9.9	    &	6.4	  &  9.1\\
\cline{1-7}
          &		&	      &	[SII] BPT &		&	      &     \\ 
\cline{2-7}
SF/Composite &	9.4	&	18.0	&	30.4 &	29.1	&	42.9  & 59.7\\
Sy2 L+U   &	82.3&	72.5	&	53.2 &	54.3	&	47.9  & 30.4\\
LINERs	  &  8.3&	9.5	    &	16.4 &	16.6	&	9.2   &  9.9\\
No. Count & 677	&	846	    &	1196 &	676	    &	438	  & 264 \\	
\cline{1-7}
          &		&	      &[OI] BPT		 &		    &	      &     \\  
\cline{2-7}
SF/Composite &4.5	&	12.2&	21.7 &	23.2	&	35.7  & 54.7\\
Sy2 L+U   & 85.3&	79.8	&	62.6 &	61.5	&	54.8  & 36.0\\
LINERs	  & 10.2&	8.0	    &	15.7 &	15.3	&	9.5   &  9.3\\
\cline{1-7}
          &		&	      &Ke06		 &		    &	      &     \\
\cline{2-7}
SF/Composite &	2.3	&	8.6	    &	15.7 &	19.0	&	32.2  & 54.2\\
Sy2 L+U   &	90.0&	83.0	&	67.4 &	64.4	&	58.5  & 37.7\\
LINERs	  & 7.7 &	 8.4    &	16.9 &	16.6	&	 9.3  &  8.2\\
No. Count$^*$ & 637	&	813	    &	1050 &	652	    &	431	  &  247	\\
\enddata											
\tablecomments{Moving from the Sy 1.0 to the Sy 1.9, each of the sub-groups consistently shift towards having larger fractions in the Composite and SF regions. Approximately $ \ge 85\%$ of Sy 1.0 lie in the Sy 2 region, while $ \leq 35\%$ of the 1.9 are remaining the the same region. In [SII], [OI], and Ke06 diagrams there is more overlap between SF and Composite, so these regions are combined, and the same with Sy 2 Upper and Lower categories. \\
$*$ [OI] emission line has fewer data points than our [SII] and [NII].
}\label{table:2}
\end{deluxetable}

\begin{figure*}[ht]         
\includegraphics[width=\textwidth]{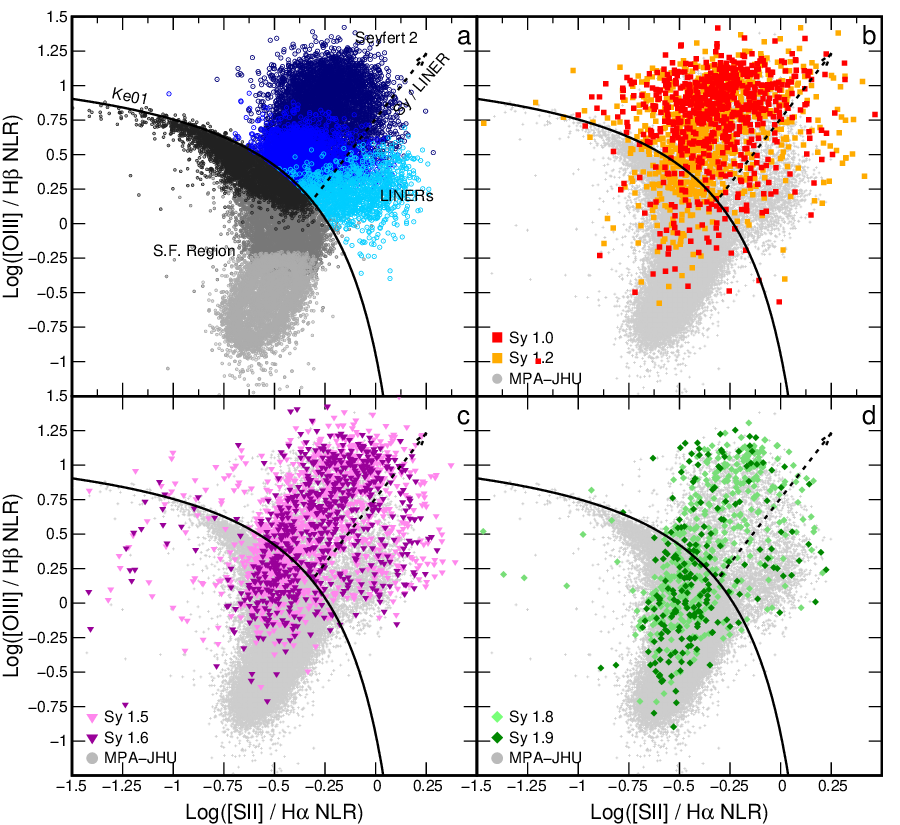}
\caption{BPT Diagrams with log([SII]/H$\alpha$ on the x-axis. The color scheme for data points is the same as in Figure \ref{fig:fig2}. 
Galaxies below the short-dash line and to the right of the Ke01 line are LINER-dominated.
(a) As before, the Star-Forming galaxy spectra are separated into three sections based on their metallicity.
 (b) More than 85\% of the Sy 1.0 and 77\% of the Sy 1.2 are in the Sy 2 ``pure AGN" region. We call these the ``strong Seyfert 1's".
 (c) The Sy 1.5 and Sy 1.6 galaxies. 30\% are in the SF region.
 (d) ``Weak Seyfert 1's": 40\% of the Sy 1.8 and 60\% of the Sy 1.9  are in the SF region. }\label{fig:fig3}
\end{figure*}

\begin{figure*}[ht]         
\includegraphics[width=\textwidth]{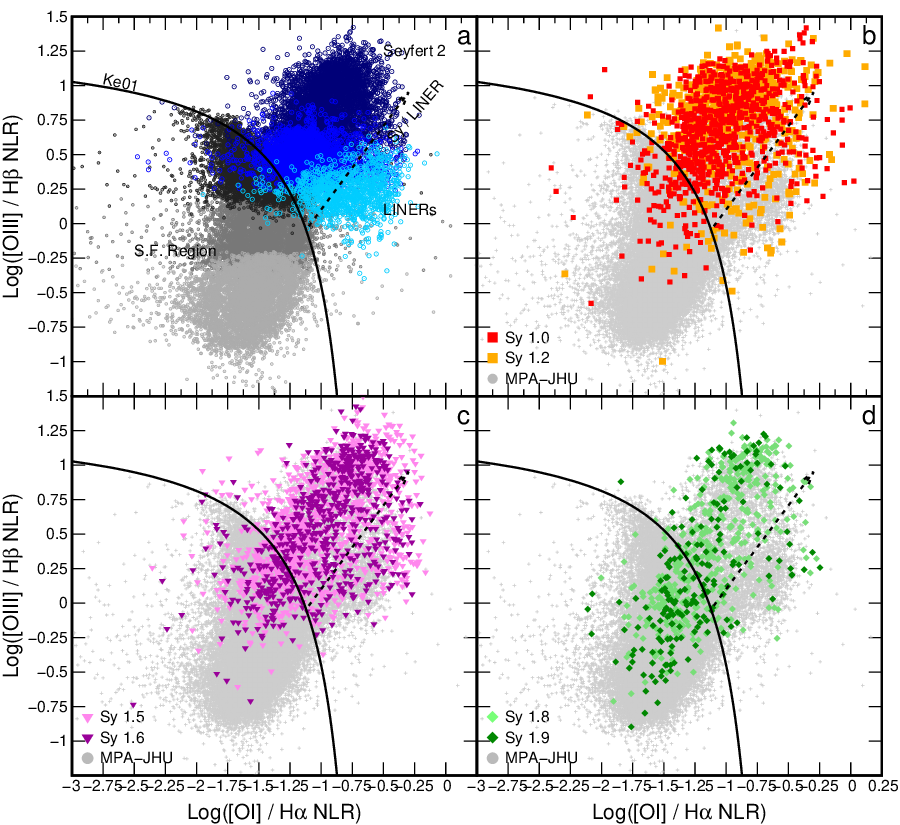}
\caption{BPT Diagram with log([OI]/H$\alpha$ on the x-axis. The color scheme as in Figure \ref{fig:fig2}.
(a)  As before, the Star-Forming galaxy spectra are separated into three (somewhat overlapping) sections.
 (b) 95\% of Sy 1.0 and 88\% of Sy 1.2 are in the AGN region.
 (c) 77\% of the Sy 1.5 and Sy 1.6 are in the AGN region.
 (d) 65\% of the Sy 1.8 NLR, and 45\% of the Sy 1.9 are in the AGN region.}
 \label{fig:fig4}
\end{figure*}

\subsection{Shifts of Seyfert 1 sub-types in BPT Diagrams Caused by HII Regions}\label{section3.1}
\begin{figure*}[ht]         
\includegraphics[width=\textwidth]{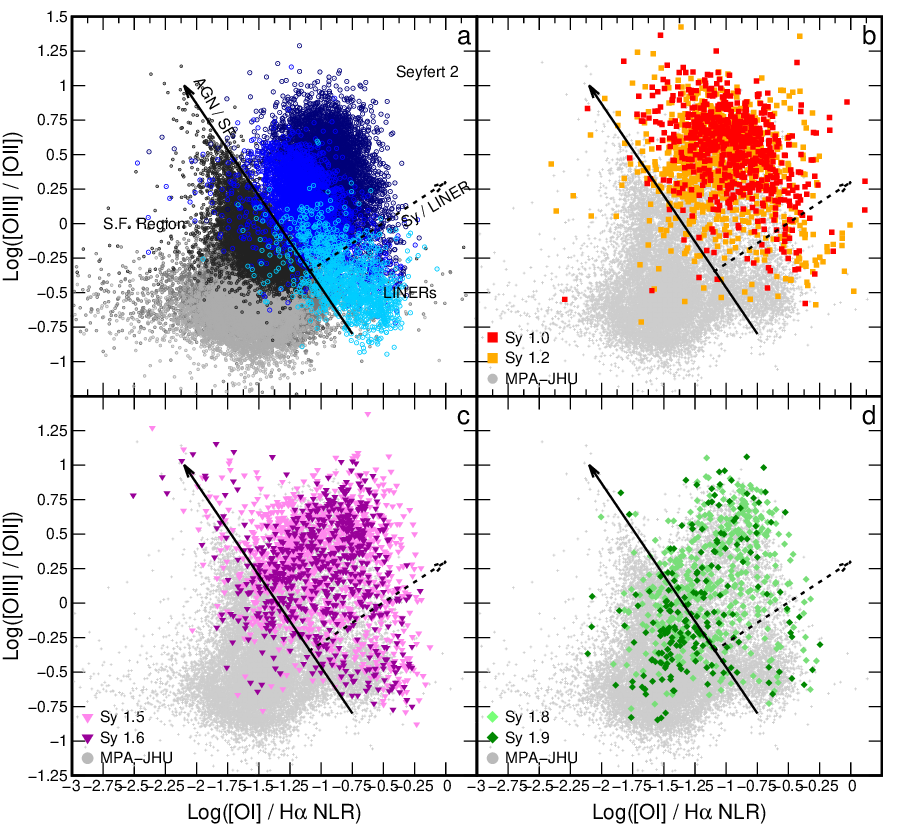}
\caption{The Ke06 diagram, of log([OIII]$\lambda$5007/[OII]$\lambda\lambda$3726+3729) $versus$ log([OI]/H$\alpha$), adopted from Kewley et al. (2006) to separate AGN, Liners and Star-forming galaxies. The solid black line (AGN/SF) separates Star-forming galaxies and AGNs. The dashed line (Sy/LINER) indicates the demarcation between Sy 2 Galaxies and LINERs. The color scheme is as in Figure \ref{fig:fig2}. (a) LINERs, Sy 2-Upper and Lower, and star-forming galaxies are plotted with different colors. $\sim7$\% of the Sy 2-Lower are found in the LINER region, while $\sim21$\% of the LINERs are above the Sy/LINER boundary.
 (b) Sy 1.0 and Sy 1.2 NLR. We see the same trend as in the traditional BPT diagrams, that they occupy the same region as the Sy 2 NLR galaxies.
 (c) Sy 1.5 and Sy 1.6 NLR. There is a small downward and rightward shift as in the other BPT diagrams. There is also a significant rightward shift into the LINER-dominated region.
 (d) Sy 1.8 and Sy 1.9 NLR. As with the BPT diagrams having log([OIII]/H$\beta$) on the y-axis, the Sy 1.9 are shifted substantially down towards the star-forming region.
}\label{fig:fig5}
\end{figure*}

The line ratio classification diagrams in Figures \ref{fig:fig2}b - \ref{fig:fig5}b all show that the BLR-dominated strong Seyfert 1 (Sy 1.0 and 1.2) spectra do indeed have NLRs overlapping with the NLRs seen in the Seyfert 2's. In the [NII] diagram 86\% and 77\%  of the Sy1.0 and Sy1.2, respectively, are found in the region of the Sy2U's (with high [OIII]/H$\beta$ ratios, indicative of {\it pure} AGN NLR emission dominating the line emission) (Table \ref{table:2}). 

However, as the relative strength of the broad H$\alpha$ component decreases, we see a growing proportion of Sy1.5's, 1.6's and-- especially--1.8's and 1.9's are {\it not} explained by the usual NLR found in Seyfert 2's. As can be seen by visual inspection of the standard BPT diagrams in Figures \ref{fig:fig2} - \ref{fig:fig4} and also in the modified Ke06  diagram (Figure \ref{fig:fig5}), compared with the NLR-dominated spectra, many of these galaxies are systematically shifted downward towards lower [OIII]/H$\beta$. While only 1.8\% of the Sy1.0s are in the SF region in the [NII] BPT diagram, fully 26.9\% of the Sy1.9s are found there, and 34.8\% are in the composite region--see Table \ref{table:2}. These significant shifts show a steadily growing contribution from HII regions in the Seyfert 1's with weaker BLRs.  

At the typical redshift of our SDSS Seyfert galaxies, the SDSS 
observing fiber (3 arcsec) captures the inner-radius$ < $2.7 kpc light from the galaxy. Since this includes not just the AGN, but also a substantial contribution of light from the host galaxy, significant emission line contamination is likely to occur whenever more than the inner Kpc of the galaxy is included in the spectroscopic aperture \citep{2018ApJ...869..138X, 2025ApJ...980...64R}.  We note that this host galaxy contamination is far larger in our new SDSS sample, compared to all previously studied Seyfert samples, because the latter were typically at several times lower redshifts \citep{1987ApJ...321..233E}. But it does not explain the systematic shift in each of the Sy1.n sub-groups, since they all have the same average redshifts (Table \ref{table:1}).

In the following section, we provide a simple two-component model which quantitatively reproduces these observational results.

\subsection{Mixing Narrow Lines from Seyfert Nuclei and Star Formation: A simple model}\label{section3.2}
We construct a simple {\it ``toy model"} that has only two emission-line components: HII-regions in typical star-forming galaxies, and the NLR emissions from gas photo-ionized by an AGN.

The possibility that the spectrum of a Seyfert 1 galaxy could be altered by the inclusion of substantial (contaminating) emission from the host galaxy, if larger physical apertures of the galaxy were observed--was quantitatively measured with 2D spectroscopic mapping of nearby Seyfert galaxies \citep{2018ApJ...869..138X, 2016ApJ...822...45T}.
The aperture-dependent spectral changes they observed are usually subtle, but our SDSS Seyfert 1 galaxies have several times higher redshifts. 
Thus the same effects they sometimes saw are evidently present even with the SDSS spectroscopy in a 3\arcsec\ fiber.
In our far larger sample we also see evidence in the BPT diagrams that many Seyfert 1 galaxies have shifted, to lower [OIII]/H$\beta$ ratios because of the inclusion of host galaxy (non-AGN) contributions. 

We are well aware that not all AGN spectra can be described by a single set of integrated line ratios, since these depend on details of the NLR such as density, location, geometry and ionizing spectral shape. Similarly, we are making a great over-simplification by describing all SF in one single integrated set of emission line ratios. The most likely deficiency in our toy model is that since it describes {\it averages} of many spectra, it fails to treat individual differences between galaxies, such as variations in heavy-element abundances \citep{2013MNRAS.431..836S}, for example.

\subsubsection{Model Procedure}\label{section3.2.1}
We begin by selecting the two model end-points, where either a pure AGN or pure HII-region spectrum produces 100\% of the H$\alpha$ line-emission.  To preserve the simplicity of our ``toy model" we combine only a single AGN spectrum with a single SF spectrum. More realistically, there must be an intrinsic variation in these two components. The most important additional parameter is the metallicity of the HII region spectrum that we selected.

{\bf Local Seyfert Nuclei avoid metal-poor host galaxies.}
Figure \ref{fig:fig2} shows the main reason that the [NII]/H$\alpha$ diagram is so effective at separating Seyfert from SF narrow lines: when Seyfert emission lines are mixed in with those of HII regions in the host galaxy, the resulting combination does {\it not} greatly decrease the total observed [NII]/$/H\alpha$ ratio.  The BPT diagram works well in the presence of this diluting ``contamination" from HII regions because those HII regions, themselves, are relatively metal- rich.  This conclusion is further supported by the recent work of \cite{2023ApJ...954..175Z, 2024ApJ...977..187Z}, who use updated photoionization models to show that the high average value of [NII]/$/H\alpha$ we measure in the average Seyfert NLR--0.9-- corresponds to a log[O/H]+12 metallicity of roughly 9.0 in the Sefyert nucleus. This agrees with the relatively high metallicity in the surrounding HII regions implied by their relatively high NII$/H\alpha$ ratios. The [NII] BPT diagram maintains a clean AGN/SF separation because AGN are almost {\it never found} in host galaxies with low metal abundances (our SF-Upper). As we showed above, this conclusion is further supported by our observation that the AGN host galaxies are massive.  The metal-poor galaxies that active nuclei avoid are the low-mass galaxies (log $M_* \sim 9$) with small or insignificant bulges, which are associated with small or insignificant central black holes, and these are much less likely to produce Seyfert nuclei \citep{2015ApJ...809...20B,2015ApJ...799..164P,2021ApJ...921...36B,2025ApJ...978..115W}. This holds because  of the well known correlation that more massive galaxies are more metal-rich \citep{2016ApJ...828...67L,2021ApJ...919..143H,2021MNRAS.501.2231S}. 
Indeed \citep{Groves2006} found that only 40 out of 23,000 Sy2 galaxies in SDSS have clearly sub-solar NLR metallicities. 
\footnote{ \cite{Juneau2014} found tentative evidence that the Seyfert galaxy NLR at high redshifts ($z \sim 2$) has lower gas metallicity--smaller [NII]/H$\alpha$ ratios--than in local Seyferts.  However, their sample was very small. While measuring this possible redshift evolution is beyond the scope of this paper, we note that if present, it will make the standard BPT classifications to separate AGN from SF galaxies more ambiguous.} 

For the SF emission component we therefore take the mid-point values of the line-ratios between SF-Lower and SF-Middle as having 0\% emissions from AGNs and 100\% from HII-regions.  For the pure AGN component, we select the median in the upper central part of the Sy2U region in all BPT diagrams (see section \ref{section2.4}). 

Since the BPT plots use logarithmic axes, we first converted line ratios to linear values. We adopt average values of H$\alpha$/H$\beta$ = 4.5 in the SF galaxies and H$\alpha$/H$\beta$ = 4.0 in AGNs. In the Ke06 diagram, We have adopted an [OII]/H$\alpha$ ratio of 1.0, which is the average in our Seyfert spectra. Because the H$\alpha$ NLR components are present in all BPT diagrams, we estimate the HII and AGN components with respect to this line. The following outlines the procedure for our {\it Toy Model}:\\

\noindent 1) Let $(x_l, y_l)$, $(x_h, y_h)$ be the $low$ (0\% AGN) and $high$ (100\% AGN) end-points, respectively.\\
    2) Let H$\alpha_{l(i)} = \{1.0, 0.99, 0.98, ..., 0.0\}$ be the set of fraction of H$\alpha$ emitted from SF galaxies and HII regions.\\
    3) Let  H$\alpha_{h(i)} = \{0.0, 0.01,0.02,0.03..., 1.0\}$ be the set of fraction of H$\alpha$ emitted from AGNs. \\
For the traditional BPT diagrams, we start from H$\alpha$ = 1.0 in the SF region (i.e. originating entirely in HII-regions):
\begin{equation}\label{eq:8}
     \text{[Line]}_i  = x_lH\alpha_{l(i)},\, H \beta_i = \frac{H\alpha_{l(i)}}{4.5},\, \text{[OIII]}_i=y_lH\beta_i
\end{equation}
and starting from H$\alpha$ = 0.0 in the SF region if all emission originates from an AGN:
\begin{equation}\label{eq:9}
     \text{[Line]}_i  = x_hH\alpha_{h(i)},\, H \beta_i = \frac{H\alpha_{h(i)}}{4.0},\, \text{[OIII]}_i=y_hH \beta_i.
\end{equation}

\begin{figure*}[ht!]  
\includegraphics[width=0.48\textwidth]{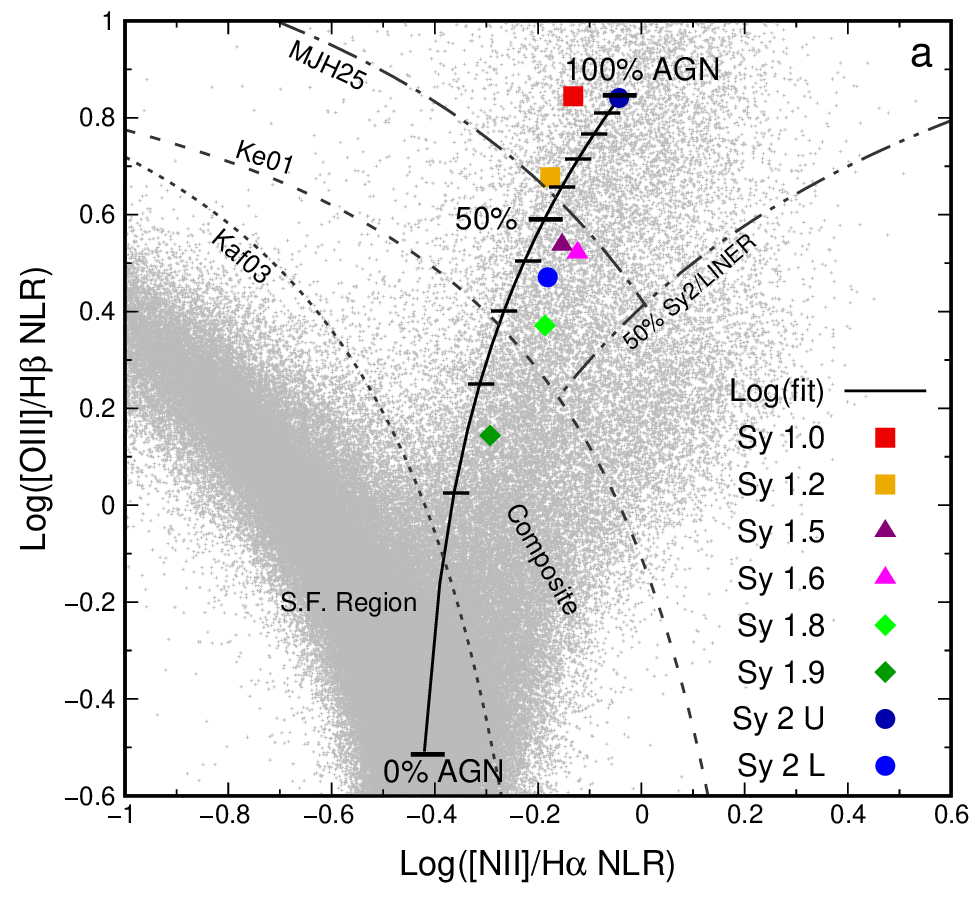}
\includegraphics[width=0.48\textwidth]{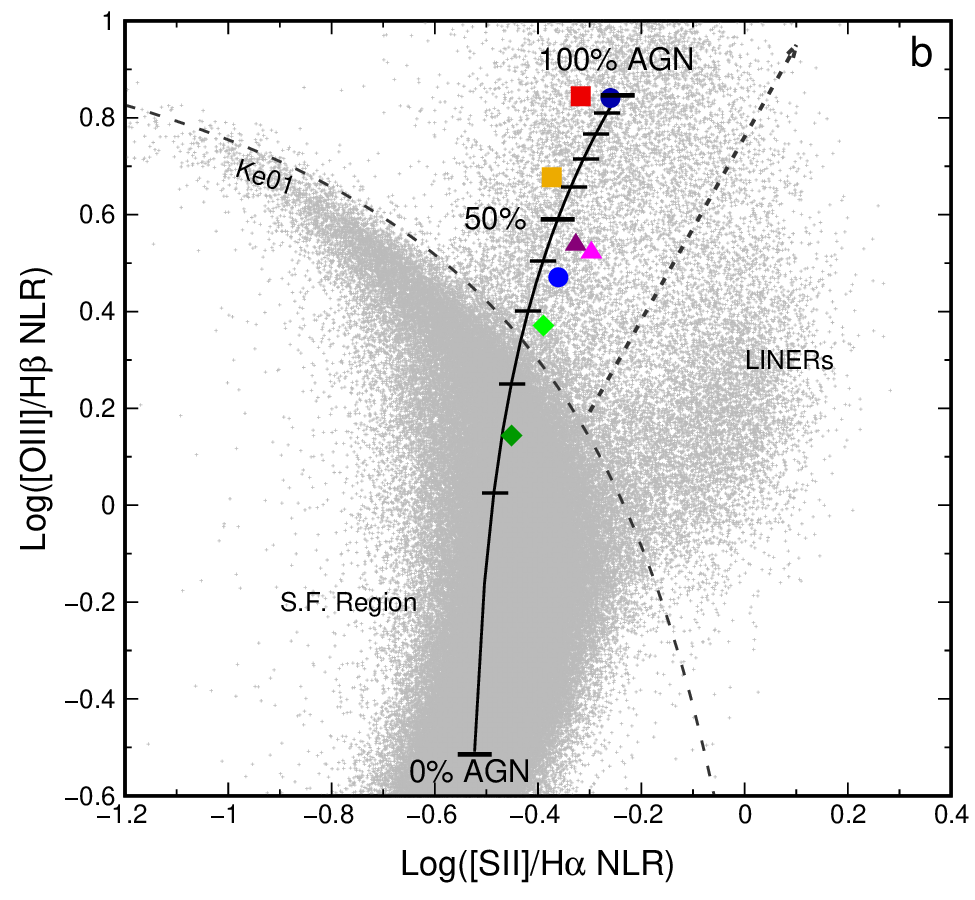}\\
\includegraphics[width=0.48\textwidth]{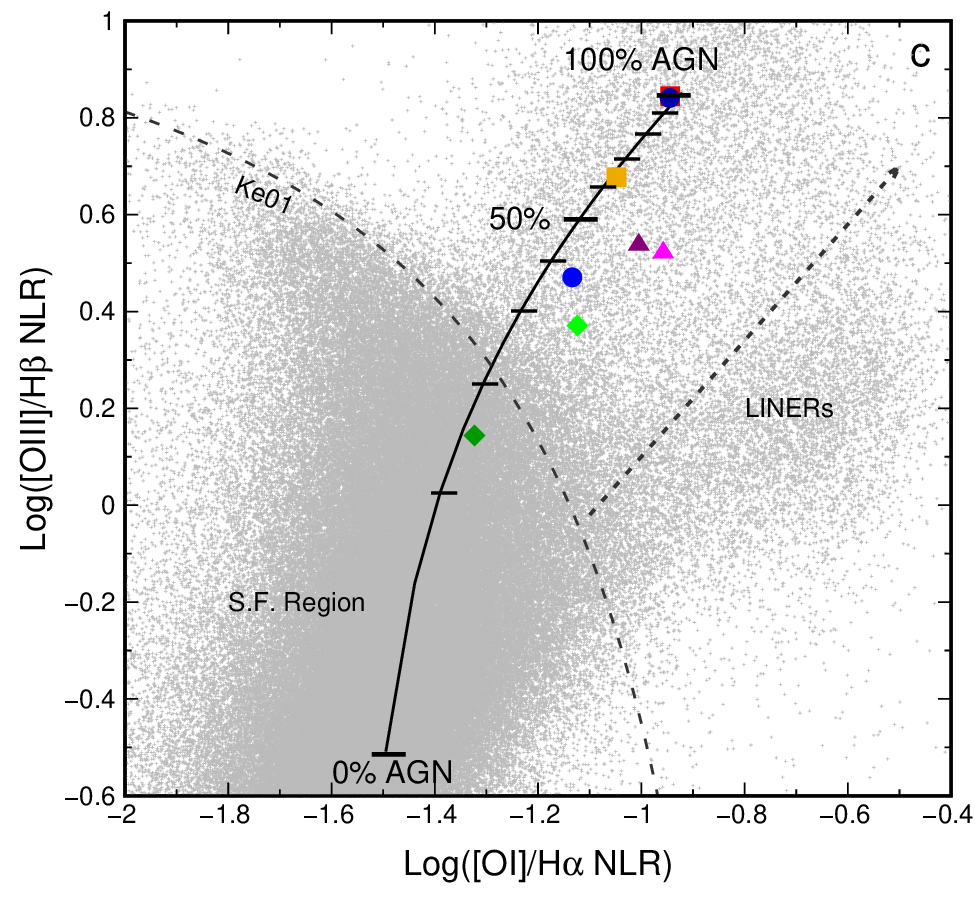}
\includegraphics[width=0.48\textwidth]{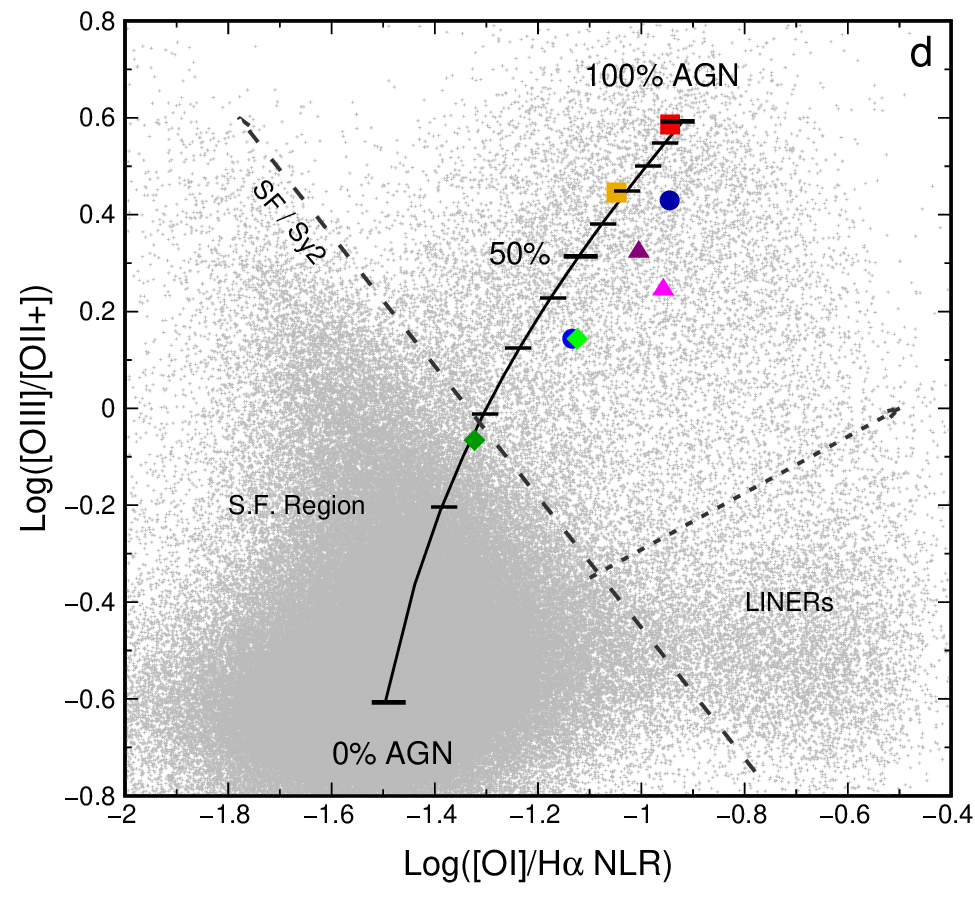}
\caption{Diagrams comparing AGN and HII region line emission with our Toy Model predictions (solid line with tic marks, see section \ref{section3.2.1}). The panel follows the same color scheme as in Figures \ref{fig:fig2}-\ref{fig:fig5}. The tics on the solid line in each graph indicate 10\% increases in H$\alpha$ contribution, from 0\% AGN activity at the bottom to 100\% at the top. The colored symbols show the medians of each Sy category. 
The strongest result is that the Sy 1.9 medians are all below the AGN/SF demarcation in each graph, including in (d) where the [OII] doublet has replaced the H$\beta$ NLR line.}\label{fig:fig6}
\end{figure*}

We follow the same procedure with the Ke06 diagram, except that we replace H$\alpha_i$ with $[OII]_i $. The resulting AGN/SF H$\alpha$ mixing line is displayed in Figure \ref{fig:fig6}. Our AGN/HII model is quite similar to the Seyfert/HII ``mixing ridge" proposed by \cite{2005MNRAS.360..565S}, who also used only the H$\alpha$ emission to classify their Seyfert galaxies.

The locus of the toy model in the [NII] BPT diagram is described by this curve:
\begin{equation}\label{eq:10}
    log\left(\frac{[OIII]}{H\beta}\right)=1.16log\left(\frac{[NII]}{H\alpha} +0.40\right) + 1.39
\end{equation}
According to our 2-component model, 45\% of the total narrow H$\alpha$ emission in the average Sy1.5 is due to HII-regions, while 55\% comes from an AGN NLR. As can be seen in Table \ref{table:3}, our toy model gives consistent results across the three BPT and Ke06 diagrams. For example the Sy 1.0's have AGN contributions to their H$\alpha$ lines of 88\%, 91\%, 95\%, and 95\% based on the [NII], [SII], [OI] BPTs, and Ke06 diagrams respectively. For the Sy 1.9's the corresponding AGN contributions to H$\alpha$ are 15\%, 14\%, 14\%, and 16\% for the same four classification diagrams. The full results for all Sy1.n sub-categories and Sy2L and Sy2U are given in Table \ref{table:3}. Scanning across each row we see that for a given Seyfert type, the AGN and SF fractions determined from the 4 different line ratios agree to within a few percent. This good consistency supports the view that our toy model is indeed capturing valid information about the amount of SF contributions to the spectra.

According to our toy model, our boundary between Seyfert 2U and Sy 2L corresponds to 56\%/44\% of the H$\alpha$ flux produced by AGN/H II regions.
\cite{2013MNRAS.431..836S} analyzed the narrow emission lines in a smaller sample of SDSS spectra of Seyfert 1 galaxies (which included detections of lines with lower SNRs). 
The locations of the narrow emission line ratios in BPT diagrams (their figure 2) show similar results to ours: They classified about 20\% of their Seyfert 1's as AGN/Star formation composites, with a small (5\%) contamination from LINERs.
They further subdivided their galaxies by the luminosity of their broad H$\alpha$ lines.  This corresponds crudely to our classification into Seyfert 1 sub-categories based on the ratio of their broad to narrow H$\alpha$ line components. Using the broad H$\alpha$ luminosity gave them a similar result to what we found. The first 3 rows of their figure 3, with increasing L(Broad H$\alpha$) resemble the BPT diagrams we found for Seyfert 1.9+1.8, Seyfert 1.6+1.5, and Seyfert 1.2+1.0, respectively. We both find as the broad component becomes relatively weaker, the narrow line classification shifts towards a higher proportion of star formation diluting the observed line ratios. This is consistent with our results, allowing for the systematic difference in our samples--the \cite{2013MNRAS.431..836S} sample contains a higher proportion of more luminous AGN, and a correspondingly lower proportion of weak Seyfert 1's (1.8's+1.9's) than ours. 

A similar two-component theoretical photoionization model was applied to 2766 Seyfert 2 galaxy spectra from SDSS by \cite{Thomas2018}. They also found that major fractions of the narrow emission lines are produced by HII regions, especially for those Seyfert galaxies with lower values of log [OIII]/H$\beta$ ($\leq$ 0.6).
\footnote{ We caution that SF components in Sy galaxy spectra cannot be unambiguously measured from consideration of a single emission line.  For example, \cite{Kim2006} measured [OII]3727 emission in Sy1 spectra, and found that with enough lowering of the ionization parameter in a single-zone model, they could reproduce the [OII] emission without an SF component. But they could not rule out the possibility of significant SF contributions.}

\begin{deluxetable}{ccccccccc}	
\tablenum{3}					
\tablecaption{Estimated AGN and HII Region Contributions to the H$\alpha$ emission.}	 
\tablewidth{0pt}					
\tablehead	
 {					
\colhead{Galaxy} &
\multicolumn{2}{c}{Log([NII]/H$\alpha $)} &
\multicolumn{2}{c}{Log([SII]/H$\alpha $)} &	            
\multicolumn{2}{c}{Log([OI]/H$\alpha $)} & 
\multicolumn{2}{c}{Log([OIII]/[OII])}\\   
\cline{2-9}
\colhead{Type} &
\colhead{AGN \%} &	\colhead{SF \%}	&	
\colhead{AGN \%} &	\colhead{SF \%}	&
\colhead{AGN \%} &	\colhead{SF \%}	&
\colhead{AGN \%} &	\colhead{SF \%}            
}					
\startdata					
Sy 1.0	&	88	&	12	&	91	&	9	&	95	&	5	&	94	&	6	\\
Sy 1.2	&	60	&	40	&	60	&	40	&	62	&	38	&	66	&	34	\\
Sy 1.5	&	45	&	55	&	46	&	54	&	48	&	52	&	57	&	43	\\
Sy 1.6	&	45	&	55	&	45	&	55	&	50	&	50	&	54	&	46	\\
Sy 1.8	&	30	&	70	&	28	&	72	&	29	&	71	&	35	&	65	\\
Sy 1.9	&	15	&	85	&	14	&	86	&	14	&	86	&	16	&	84	\\
Sy 2 L	&	38	&	62	&	37	&	63	&	37	&	63	&	34	&	66	\\
Sy 2 U	&	98	&	2	&	95	&	5	&	95	&	5	&	74	&	26  \\
Ke01  &	50	&	50	&	70	&	30	&	80	&	20	&	79	&	21  \\
Kaf03 &	67	&	33	&	...	&	...	&	...	&	...	&	...	&	...  \\
\enddata											
\tablecomments{Average Estimated percentage of AGN and HII Regions contribution in each galaxy group for our Toy Model. This simplified analysis excludes any possible LINER contribution (see section \S \ref{section3.3}). The top column headers indicate the x-axis on the BPT diagrams, except for the right-most column header, which indicates the y-axis as adopted from \cite{2006MNRAS.372..961K}. Ke01 and Kaf03 indicate the percentage of the type of galaxies laying above or below these boundary lines in the BPT diagrams.}\label{table:3}			
\end{deluxetable}

\subsection{LINER Components}\label{section3.3}
Besides overlooking variations in the metallicity of HII regions, the other main limitation of our simple toy model is its omission of a third possible source of emission lines in galaxy spectra: ``LINER" spectra are empirically defined as having larger [NII]/H$\alpha$ ratios than SF galaxies, but lower [OIII]/H$\beta$ than classic Seyferts \citep{1980A&A....87..152H}. The inability of SF mixing to explain all the data is a weakness of earlier (over-simplified) scenarios in the literature which was remedied in our earlier analysis of these diagrams (M17). Using a cruder methodology, we found a small contribution of LINER emission in a small sample of Seyfert galaxy spectra.

The physical explanation for LINER emission lines is still under debate.  Some researchers argue that most LINER spectra are produced by photoionization by an AGN continuum, but their gas is much further out from a relatively fainter nucleus, making its ionization parameter an order of magnitude lower than it is in the classic Seyfert 2 NLR \citep{1983ApJ...264..105F, 2013ApJ...763..145D}.
On the other hand, other researchers argue that the LINER spectrum is mainly produced by other galactic phenomena not directly connected to an AGN, such as winds and shocks in the ISM \citep{2000ApJ...529..219S}\footnote{
LINER-like emission can be seen in the Diffuse Interstellar Medium of galaxies \citep{2017ApJ...850..136S}. Like the Diffuse Ionized Gas (DIG) seen at high latitudes in our Milky Way, the DIG produces strong [NII] and [SII] emission lines, but not [OI], in contrast to what we see in many of our Seyfert 1 spectra.}.

In all of the emission line ratio diagrams there is a continuous variation of observed galaxy spectra from what are classified as Seyfert spectra to LINERs. 
In Table \ref{table:2}, we give the percentages of each sub-group of Seyfert galaxies that fall below the Seyfert 2/LINER 50\% mixing line of M17.  
The percentages of each Seyfert 1 sub-category that have a strong enough [OI] emission to classify them as LINERs are also given in Table \ref{table:2}. 
\footnote {Of the three diagnostic BPT diagrams, the one using [NII] provides the sharper restriction against LINERs. In other words, galaxies are somewhat more likely to satisfy the LINER conditions in the [SII] and [OI] BPT diagrams, which have more inclusive LINER definitions.}
Across all Sy 1.0, 1.2, 1.8, and 1.9 this fraction is on average 8 percent. This shows a strong empirical fact that the two Sy groups with strong broad permitted lines are almost always associated with the NLR, and rarely associated with LINER emission. But the relatively small LINER contamination in the Seyfert 1.8's and 1.9's is probably because their spectra have such strong line emission from star-formation that this overwhelms a relatively weak LINER contribution.

However, for the {\it intermediate} Seyfert 1's, ie., the 1.5s and 1.6's, we see a quite significant increased percentage: on average 15\% of them lie within the LINER-dominated region in all of these diagnostic diagrams. The simplest explanation is that the LINER contribution is very weak and is easily overwhelmed by other components when present: the strong AGN NLR in strong Seyfert 1's (1.0's and 1.2's), and the strong star formation in weak Seyfert 1's (1.8's and 1.9's).  It is the intermediate Seyfert 1's (1.5's and 1.6's) where these other components are both weak enough that the relatively weak LINER contribution is observationally most evident. 

The AGN/SF [NII] BPT diagram shows a strip of galaxies with line ratios
falling between the SF boundary and the pure Seyfert boundary lines
established by \cite{2001ApJ...556..121K} (Ke01-line in Figures \ref{fig:fig2}, \ref{fig:fig3}, and \ref{fig:fig4}) and \cite{2003MNRAS.346.1055K} (Kaf03-line in Figure \ref{fig:fig2}). 
We describe galaxies in this intermediate strip as having ``Composite" spectra. Although many of these composite spectra are indeed explained by
a simple pure NLR-plus-HII mix, roughly half of all these ``composite" AGN have narrow emission lines better described as having a predominantly LINER rather than a Seyfert 2 nuclear component.
We agree with the conclusion of \cite{2021ApJ...922..156A} that the main reason for the  wide range of line ratios seen in LINERs is intrinsic variation (``cosmic scatter"), and {\it not} mixing with line emission from HII regions  \footnote{ By construction, our emission line galaxy sample has mostly LINERs that \cite{2021ApJ...922..156A} classify as ``H-LINERs", with [OI]6300 emission. We have relatively few of their ``S-LINERs", i.e. ``soft LINERS" which have small [OIII]/H$\beta$ ratios ($<1$), and whose properties overlap much more with SF galaxies}.

\subsection{Stellar Masses}\label{section3.4}

Although the detailed individual spectral modeling needed to calculate accurate stellar masses of our galaxies is beyond the scope of this work, we obtained rough estimates for each of the galaxy sub-groups in our study. We base these on the integrated i-band light given in the SDSS photometric database (https://www.sciserver.org/, using DR7 to match our spectroscopic data). Since the average redshifts of all of our AGN sub-groups are the same, it is not surprising that their average total i-band luminosities, assumed to be almost entirely starlight, are also very similar.  In fact, the average absolute magnitude of {\it all} categories of Seyfert 1's, Sy2's and LINERs are within 0.1 mag of -21.6 (Table \ref{table:4}). Assuming the mass/light ratio of typical early type galaxies, of 2.5 in solar units, this corresponds to $10^{10.7} M_{Sun}$. 
We obtain similar stellar masses for our ``Lower" and ``Middle" SF galaxies, which have mean absolute i magnitude of -21.1 and -20.8 respectively. (The average absolute magnitude of our SF ``Lower" galaxies, which we argue do not match the higher metallicities of Seyfert host galaxies, is -19.8).    

The high stellar masses of Seyfert host galaxies will motivate a simple model described below which is based on the assumption that the AGN host galaxies  are most similar to either the SF-Lower or SF-Middle samples. 
This high stellar mass of AGN host galaxies agrees with the average value found by \cite{2003MNRAS.346.1055K}. The great majority of SDSS AGN that we and others have studied have massive host galaxies (see also \cite{Reines2013} and \cite{Sartori2015}). This also further supports our claim that they are metal-rich. 

\section{Predicting [NII] Flux from [SII] to Deblend Halpha}\label{section4.0}

A large and growing body of spectroscopy of faint galaxies
is being obtained at low spectral resolution. At $\lambda / \Delta \lambda \leq 300$, these spectra are unable to separate the H$\alpha$ line emission from the [NII]$\lambda\lambda$6584+6548 doublet.
Thus, for example, low-resolution spectra cannot use H$\alpha$ luminosities, unless they are first corrected for the blended [NII] contributions (\cite{2021MNRAS.501.2231S, 2002ApJ...581..205H}).
The same problem is present in most narrow-band imaging studies of H$\alpha$ \citep{2016ApJ...822...45T, 2012ApJ...747L..16L}.
In Seyfert 1 spectra with broad H$\alpha$ lines, the [NII] lines may be difficult to separate even using higher resolution spectroscopy (\cite{1982ApJ...256...75L}). A practical solution, put forward by M17, was to measure the nearby isolated [SII] doublet to predict what the strength of the blended [NII] lines should be. In their section \S 3.3 they proposed that a ``universal" reddening-independent average value of log([SII]$\lambda\lambda$6716+6731)/[NII]6584)) = $-0.23 \pm 0.18$ could be used for all types of AGN.

\begin{figure}[ht!]
\includegraphics[width=0.9\textwidth]{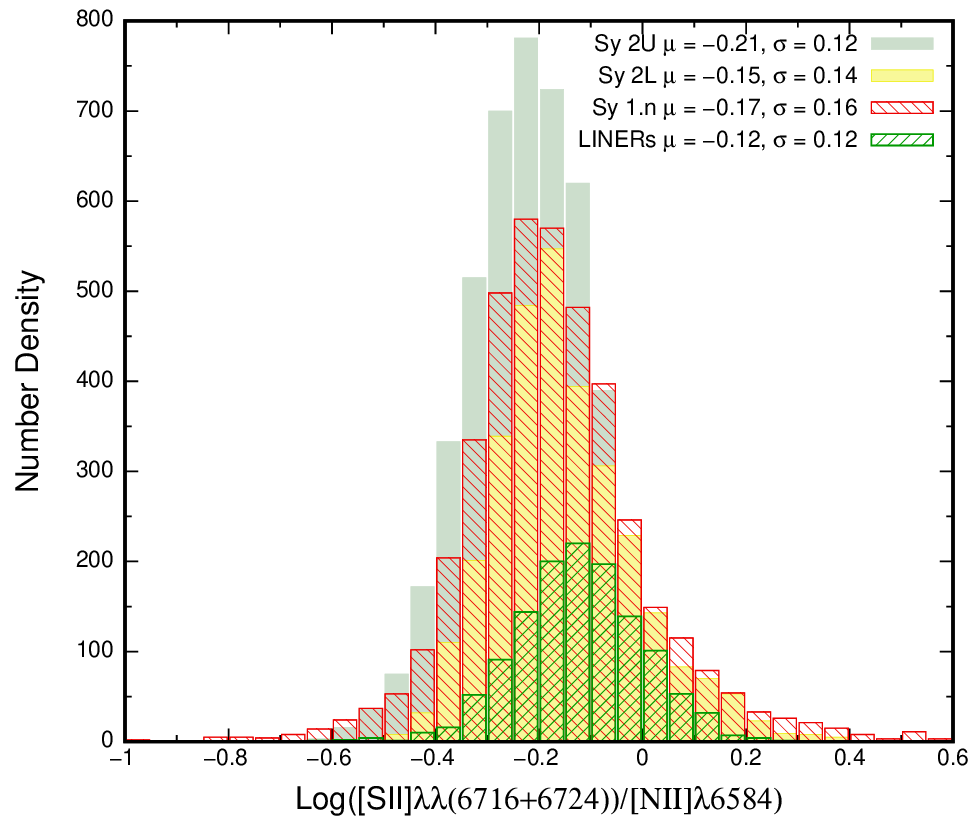}
\caption{Histogram of observed ratios of [SII] doublet to [NII]6584 emission line.  This shows that for a measured [SII] doublet flux, one can predict the [NII]6584 line flux, with an one-sigma accuracy of 30\%. However, the correct ratio varies slightly from Seyfert 2, to Seyfert 1, to LINER galaxy spectra, as shown by the mean values. \label{fig:fig7}}
\end{figure}

\begin{deluxetable*}{lcc}	
\tabletypesize{\normalsize}
\tablenum{4}					
\tablecaption{Absolute Magninute i-band}	
\tablewidth{1000pt}
\tablehead					
{					
\colhead{Galaxy Type} & \colhead{Log([SII]($\lambda\lambda6717+6731$)/[NII]))} & \colhead{AbsM i}	
}					
\startdata														
Sy 1.0	&	-0.18	$\pm$	0.18	&	-21.77	$\pm$	0.78	\\
Sy 1.2	&	-0.19	$\pm$	0.20	&	-21.78	$\pm$	0.96	\\
Sy 1.5	&	-0.15	$\pm$	0.19	&	-21.66	$\pm$	0.78	\\
Sy 1.6	&	-0.13	$\pm$	0.17	&	-21.54	$\pm$	0.79	\\
Sy 1.8	&	-0.17	$\pm$	0.18	&	-21.65	$\pm$	0.87	\\
Sy 1.9	&	-0.21	$\pm$	0.36	&	-21.63	$\pm$	1.12	\\
Sy 2 L	&	-0.15	$\pm$	0.13	&	-21.35	$\pm$	0.79	\\
Sy 2 U	&	-0.21	$\pm$	0.12	&	-21.56	$\pm$	0.74	\\
LINERs	&	-0.12	$\pm$	0.12	&	-21.79	$\pm$	0.75	\\
SF L	&	-0.10	$\pm$	0.10	&	-21.07	$\pm$	0.79	\\
SF M	&	0.17	$\pm$	0.13	&	-19.97	$\pm$	0.90	\\
SF U	&	0.41	$\pm$	0.14	&	-18.99	$\pm$	1.08	\\
\enddata											
\tablecomments{Averages of Log([SII]$\lambda\lambda6717+6731$)/[NII])) and i-band Absolute Magnitude.}\label{table:4}
\end{deluxetable*}

Our new analysis with far larger samples of AGN spectra now allows us to refine the M17 prediction. The mean and standard deviations of log(([SII]$\lambda\lambda$6716+6731)/[NII]6584) are given in Figure \ref{fig:fig7}. 
As expected, the Sy 2U have a mean value of -0.21, almost the same as what M17 adopted. This is because both of those datasets refer to strongly ``AGN-dominated" spectra. However, the mean value of  log([SII]/[NII]) in all other Seyfert sub-classes is slightly higher: -0.17. Although that is only a 10\% increase, the samples are so large that this small offset is accurately determined.  Therefore, the original suggestion of M17 still works well with small modifications:
the [NII]$\lambda 6584$ line flux can be
predicted by multiplying the [SII]$\lambda \lambda6716+6731$ 
flux by a constant ratio of 1.48. 
The large intrinsic scatter, however, means that any individual [NII] flux will only be predicted from [SII] with an accuracy of about 30\%.
And yet a slightly different multiplicative ratio should be used for predicting [NII] in LINERs, which should be 1.32 times [SII]$\lambda \lambda 6716+6731$. Not surprisingly, this ratio is the same as observed in metal-rich star forming regions, as the host galaxies of LINERs are generally metal-rich. We did not find any correlation between the [NII]/[SII] ratio and the host galaxy luminosity.  This suggests that the ratios we observe are not impacted by metallicity variations, perhaps because these are not large in our Sy galaxy samples.

\section{Reddening Estimates from Narrow Emission Line Ratios}\label{section5.0}
M17 considered the flux ratios of several pairs of narrow emission lines at widely differing wavelengths. If the intrinsic (unreddened) ratio of a line in the red to a line in the blue is a known constant, then the observed ratio measures the gas reddening. In particular, the Balmer decrement, (H$\alpha$/H$\beta$), of the relatively low-density gas producing the narrow lines has long been used as a reddening indicator, with an assumed intrinsic value of 2.86 (Case B recombination), to 3.1 (\cite{1984PASP...96..393G}), to somewhat larger values ($\sim 3.5$ \cite{1983ApJ...264L...1M}; see also \cite{1986ApJ...310..679M, 2021ApJ...922..272Y}).

\begin{figure}[ht!]
\includegraphics[width=1.0\textwidth]{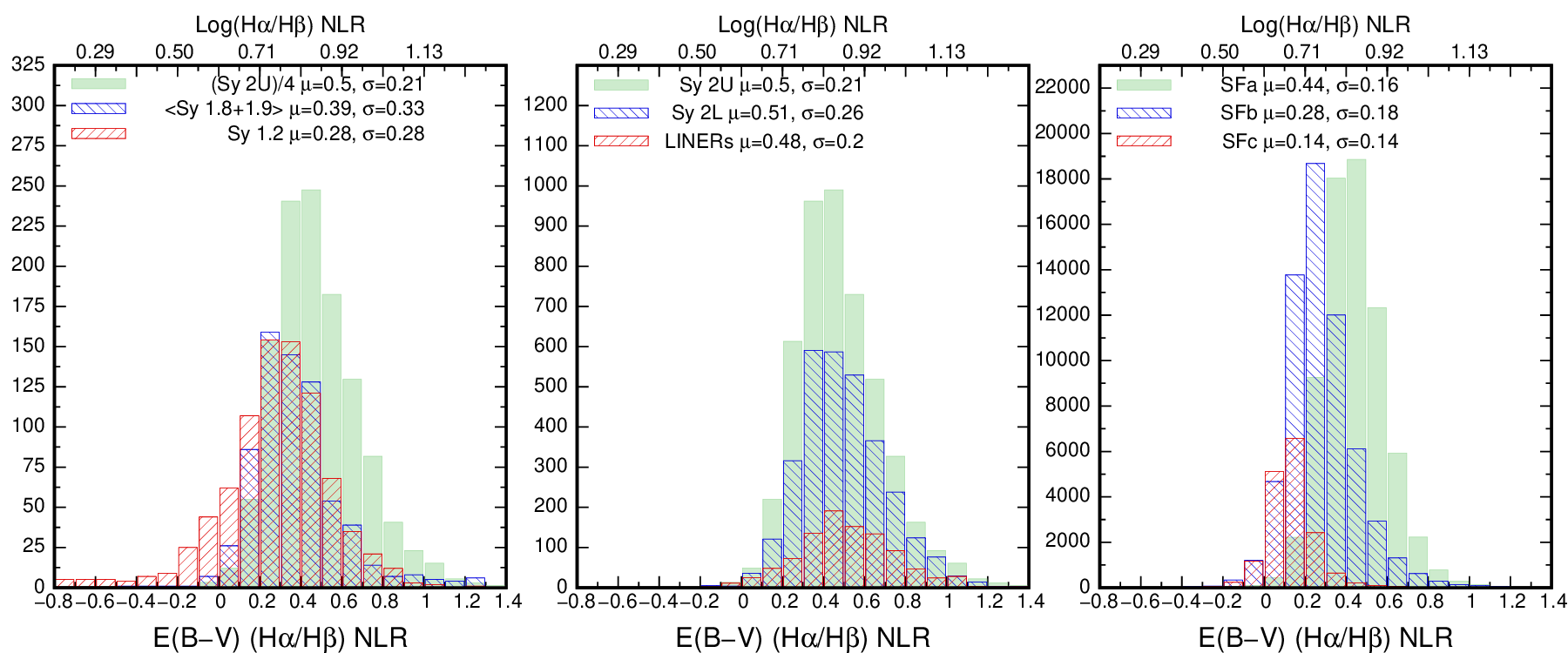}
\caption{Reddening estimates (with Number Density on the y-axis) from narrow-line Balmer decrement, assuming intrinsic H$\alpha$/H$\beta$= 2.86, and standard reddening law. Reddenings of Sy2U, Sy2L, and LINERS ({\it middle}) are larger than Sy1's ({\it top}), especially Sy1.2. The bottom panel shows SF galaxies' reddenings drop 0.3mag as their gas metallicity drops. The Sy1.8+1.9 galaxies, with contamination from star formation, have similar reddenings to SF-Lower, presumably because they all have high abundances.\label{fig:fig8}}
\end{figure}

\subsection{Milky Way Reddening}\label{section5.1}
We plotted the distributions of E(B-V) we determined from the H$\alpha$/H$\beta$ ratios as a function of galactic latitude. At high galactic latitudes (greater than 40 degrees), there is no trend for the Balmer decrement to decrease with latitude.  However, going down to $|$b$|$ = 20 degrees, log(H$\alpha$/H$\beta$) does increase subtly, by 0.03 dex. That is the closest to the Milky Way plane that our galaxy samples reach, and the corresponding excess reddening due to our Milky Way, compared with high galactic latitudes, is $\Delta$E(B-V) = 0.06 mag \citep{2013ApJ...763..145D}.  This subtle excess can be roughly approximated by the usual cosecant law for an infinite plane slab:
$\Delta$(E(B-V)) = 0.02 mag / sin(b)  .
Fortunately, the great majority of our galaxies are more than 40 degrees out of the galactic plane, {\it independent of spectral type. } Therefore Milky Way reddening has a negligible effect on the average emission line ratios we analyze, and we therefore ignore it.

\subsection{Average Reddenings of AGN Categories}\label{section5.2}
Figure \ref{fig:fig8} shows the distributions of E(B-V) determined from the H$\alpha$/H$\beta$ ratios of the narrow line components.  
Since negative reddenings are unphysical,
the reddening distributions truncate around zero reddening, except for the non-negligible observational uncertainties.
The mean reddening decreases as the relative strength of the BLR increases. As tabulated in Table \ref{table:5}, the Seyfert 2 galaxies (and LINERs) have $<E(B-V)> \sim 0.5$, while the weak Seyfert 1 (1.8 +1.9) galaxies have $<E(B-V)>=0.39$. 
Converting E(B-V) to visual extinction by multiplying by 3.1, our measures are reasonably consistent with the average values \cite{Vaona2012} found for their smaller sample of Seyfert 1's ($A_V$=0.9 mag) and Seyfert 2's ($A_V$=1.4 mag).

\begin{deluxetable*}{lrrrrrrrrrrrr}
\tablenum{5}
\tablecaption{E(B-V) from NLR H$\alpha$/H$\beta$ \label{tbl3}}
\tablewidth{700pt}
\tabletypesize{\footnotesize}
\tablehead{
\colhead{Statistical} & 
\colhead{Sy 1.0} & \colhead{Sy 1.2} & 
\colhead{Sy 1.5} & \colhead{Sy 1.6} & 
\colhead{Sy 1.8} & \colhead{Sy 1.9} & 
\colhead{Sy 2.0} & \colhead{Sy 2.0} & \colhead{LINERs} & 
\colhead{SF} & \colhead{SF } & \colhead{SF}\\
\colhead{Data} & 
\colhead{} & \colhead{} & 
\colhead{} & \colhead{} & 
\colhead{} & \colhead{} & 
\colhead{Lower} & \colhead{Upper} & \colhead{} &
\colhead{Lower} & \colhead{Middle} & \colhead{Upper}
} 
\startdata
Mean	&	0.13	&	0.28	&	0.33	&	0.36	&	0.39	&	0.40	&	0.51	&	0.50	&	0.48	&	0.44	&	0.28	&	0.14	\\
Standard Error	&	0.02	&	0.01	&	0.01	&	0.01	&	0.01	&	0.02	&	0.01	&	0.00	&	0.00	&	0.00	&	0.00	&	0.00	\\
Median	&	0.25	&	0.29	&	0.32	&	0.33	&	0.34	&	0.34	&	0.50	&	0.48	&	0.45	&	0.43	&	0.26	&	0.12	\\
First Quartile	&	0.05	&	0.14	&	0.18	&	0.22	&	0.24	&	0.22	&	0.35	&	0.35	&	0.33	&	0.34	&	0.17	&	0.07	\\
Third Quartile	&	0.42	&	0.44	&	0.44	&	0.45	&	0.47	&	0.46	&	0.67	&	0.62	&	0.60	&	0.53	&	0.36	&	0.19	\\
Standard Deviation	&	0.52	&	0.28	&	0.24	&	0.26	&	0.26	&	0.39	&	0.26	&	0.21	&	0.21	&	0.16	&	0.18	&	0.14	\\
\enddata
\tablecomments{Statistics of E(B-V) showing results for Log(H$\alpha$/H$\beta$) Sy 1.n (NLR data only), Sy 2s, LINERs, and SF galaxies. The Seyfert and LINER data are from the Hao-sample, while the SF data are from MPA-JHU DR7 database. There is a clear progressive increase in $<E(B-V)>$ values from Sy 1.0 to Sy 2U. The SF-Upper has significantly less reddening than SF-Lower, where upper and lower SF still correspond to their respective locations in the [NII] BPT diagram. For Sy 2U, LINERs, and all the SF galaxies the differences are highly significant with $\sigma/\sqrt{n} <$ 0.004.}\label{table:5}
\end{deluxetable*}

As our above analysis of line-ration diagrams indicated, the host galaxies of Seyfert galaxies most resemble either our SF-Lower or SF-Middle spiral galaxies, which have average
$<E(B-V)>$=0.44 and 0.28 mag, respectively.
If this is correct, it suggests that the reddening of AGN emission lines is similar to that of SF regions in similar galaxies.  However the scatter is large, and this overlap might be coincidental.

Our one clear result is that 
the BLR-dominated strong Seyfert 1's (Sy1.0's and Sy1.2's) have smaller
extinction-- $<E(B-V)>=0.13$ and 0.28, respectively-- than any of the other (less BLR-dominated) Seyfert galaxy classes. 
Because of the large sample sizes, these differences are statistically significant. As discussed further below, the differing narrow-line extinctions are a contradiction of the simplest AGN unification scheme, in which all Seyfert narrow line regions are the same.

One possible explanation of this reddening difference has been considered by \cite{Lagos2011}, who analyzed the axial ratios, (b/a), of the Seyfert host galaxies.  Their sample of Seyfert 2 galaxies had the same b/a distribution as a control sample of spiral galaxies without AGN. However, their Seyfert 1 galaxies were strongly biased to face-on disk orientations. To check this in our sample, we obtained the axial ratio measurements of all our galaxies in each sub-category from SDSS direct imaging catalogs in the g, r and i bands (https://www.sciserver.org/, using DR7 to match our spectroscopic selection). \footnote {The choice of wave-band does not matter, since the i-band images--more influenced by bulge light than the g-band images which are more influenced by disk light, only differ by at most 0.01 in b/a ratios.} The average (b/a) ratio in {\it all} of our Seyfert 1 sub-groups is 0.74, which is only slightly larger than the 0.69 average we find in our Seyfert 2 host galaxies. Even this minor difference may not reflect a systematic difference in inclination angle. This is because our control sample of non-AGN galaxies shows that the SF-Lower galaxies have $<b/a> = 0.72$, compared with the SF-Middle and SF-Upper galaxies with $<b/a> = 0.61$. We attribute that small difference to slightly larger bulge/disk ratios in the more massive SF-Lower galaxies. And by implication, most of the even smaller Sy1 versus Sy2 difference is probably due to slightly higher bulge/disk ratios in the Sy1's.

\subsection{Comparison of Balmer Decrements with [SII]/[OII] Ratios}\label{section5.3}
M17 pointed out that if a second reddening-sensitive line ratio is compared with the Balmer decrements, then the combination of the two line ratios of all galaxies should lie along a single reddening track.  With a large galaxy sample, the track might be expected to start near the intrinsic values of the ratios, since the least possible reddening in a galaxy is zero. The slope of two line ratios is predicted by the shape of the interstellar reddening law.

\begin{figure}[ht!]
\includegraphics[width=0.9\textwidth]{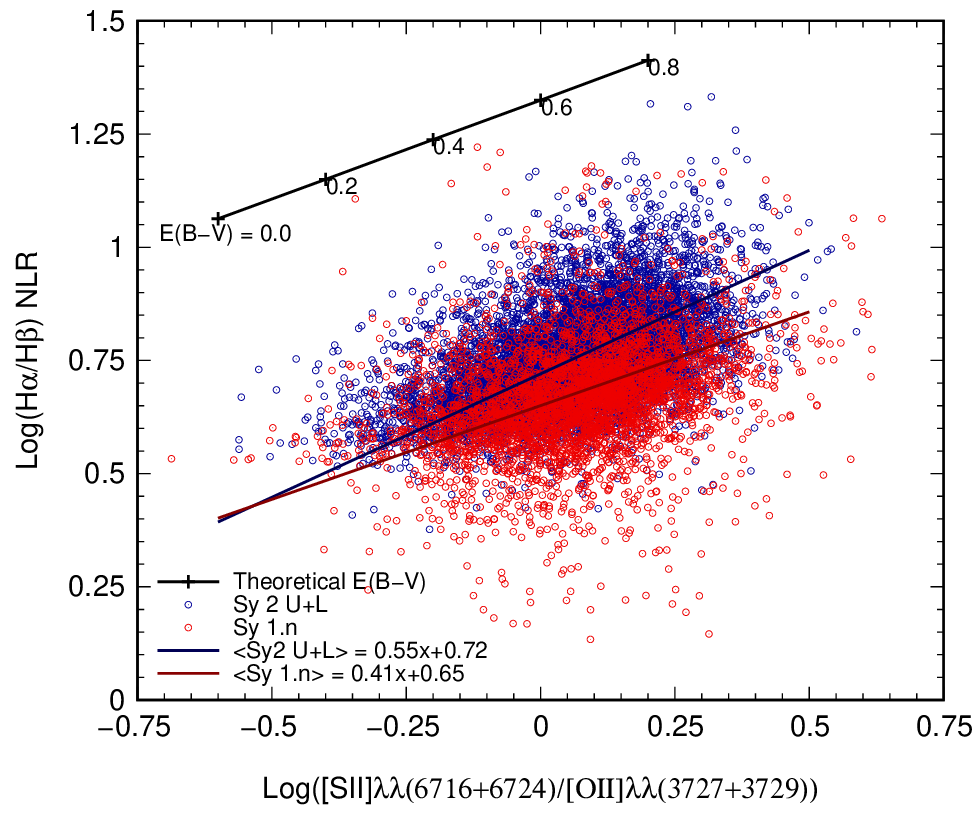}
\caption{Comparison of two proposed reddening indicators, for Seyfert 1 (red) and Seyfert 2 (blue) galaxy spectra. If both the vertical axis (H$\alpha$/H$\beta$) and the horizontal axis ([SII]/[OII] doublet ratio) measured the same gas reddening, then for a standard interstellar reddening law, all galaxy points should lie along a line with the logarithmic slope of +0.44, shown by the arbitrarily placed heavy solid black reddening line. The actual correlations (blue and red solid lines from orthogonal proper least squares fits ($C=Ax+B$) show the slope and intercepts for the Seyfert 2s data are somewhat steeper than expected (Table \ref{table:6}).}\label{fig:fig9}
\end{figure}

This led M17 to recommend the ratio of ([SII]$\lambda\lambda$6716+6731)/([OII]$\lambda\lambda$3726+3729), since both doublets are produced by singly ionized gas, so that their ratio would show small {\it intrinsic} variation.  Due to the particularly wide wavelength separation of these two doublets, their observed ratio should then be primarily sensitive to dust reddening, providing a useful supplement to the usual Balmer decrement estimator (H$\alpha$/H$\beta$, \cite{1983ApJ...265...92M}).   
Assuming fixed intrinsic line ratios subjected to uniform dust reddening screens following a standard Milky Way extinction law (\cite{1989ApJ...345..245C}), the observed line ratios plotted in Figure \ref{fig:fig9} should follow a track starting from the unreddened values, and sloping up to the right with a slope of +0.44 in log-log coordinates. 

Using our newly assembled large datasets of SDSS spectra, we have re-tested the hypothesis that reddening of the narrow-emission-line region determines both of the observed line ratios, for all different types of emission-line galaxies. Our Figure \ref{fig:fig9} is an improved version of M17's figure 2.  In Table \ref{table:6} we present the best-fit from an orthogonal proper least squares fit, using the 
{\fontfamily{qcr}\selectfont
FORTRAN
} program written and described by \cite{1996ApJ...470..706A}

This bisector method minimizes the distance from the datapoints to the regression line in both x and y dimensions, and is therefore appropriate when there are comparable uncertainties in both the x- and the y-axis,
as is the case in our Balmer/([SII]/[OII]) line-ratio correlations for each galaxy spectral class.

The scatter in the correlations is large.
The Seyfert 1 galaxies, and the Sy2L all show roughly the same  [SII]/[OII] vs. Balmer decrement correlation that M17 found (their table 8). And since the slope was similar to the expected value of +0.44, this supports their claim that both line ratios consistently measure the same NLR reddenings. The observed scatter indicates that E(B-V) in Sefyert 1's can thus be determined with a one-sigma uncertainty of  +/- 0.13 mag.

However, our large new spectral samples show that this claim is {\it not generally valid} for the narrow lines of the ``pure" Sefyert 2 galaxies, the Sy2U. The problem is that for a given Balmer decrement, Sy2U galaxies have relatively lower [SII]/[OII] ratios, particularly in the more reddened galaxies. 

\begin{deluxetable}{lccc}	
\tabletypesize{\normalsize}
\tablenum{6}					
\tablecaption{Reddening Ratios\\ Log(H$\alpha$/H$\beta$) v. Log([SII]/[OII] doublets.}	
\tablewidth{0pt}
\tablehead					
{					
\colhead{Galaxy Type} & \colhead{Slope (A)} & & \colhead{Intercept (B)}	
}					
\startdata					
Sy 1.0	&	0.38 $\pm$ 0.085	& &	0.63 $\pm$ 0.007\\
Sy 1.2	&	0.45 $\pm$ 0.063	& &	0.62 $\pm$ 0.006\\
Sy 1.5	&	0.36 $\pm$ 0.039	& &	0.66 $\pm$ 0.004\\
Sy 1.6	&	0.43 $\pm$ 0.043	& &	0.65 $\pm$ 0.004\\
Sy 1.8	&	0.45 $\pm$ 0.051	& &	0.66 $\pm$ 0.005\\
Sy 1.9	&	0.41 $\pm$ 0.051	& &	0.69 $\pm$ 0.008\\
Sy 2 L	&	0.48 $\pm$ 0.016	& &	0.73 $\pm$ 0.002\\
Sy 2 U	&	0.61 $\pm$ 0.018	& &	0.72 $\pm$ 0.002\\
LINERs	&	1.18 $\pm$ 0.102	& &	0.74 $\pm$ 0.005\\
SF L	&	0.40 $\pm$ 0.002	& &	0.63 $\pm$ 0.000\\
SF M	&	0.38 $\pm$ 0.001	& &	0.67 $\pm$ 0.000\\
SF U	&	0.25 $\pm$ 0.004	& &	0.65 $\pm$ 0.002
\enddata											
\tablecomments{Best fit Reddening Ratios computed with Proper Least Square, {\it C = Ax + B}. The average slope $<A>$ for Sy 2's (L+U)= 0.55 is significantly higher than that of Sy 1's $<A>$ = 0.41. The Sy 1's and also the star-forming Lower and Middle galaxies have slopes consistent with the expected standard interstellar reddening law.}\label{table:6}
\end{deluxetable}

\section{Electron Densities in Different Galaxy Classes}\label{section6.0}
\begin{figure}[ht!]
\includegraphics[width=0.9\textwidth]{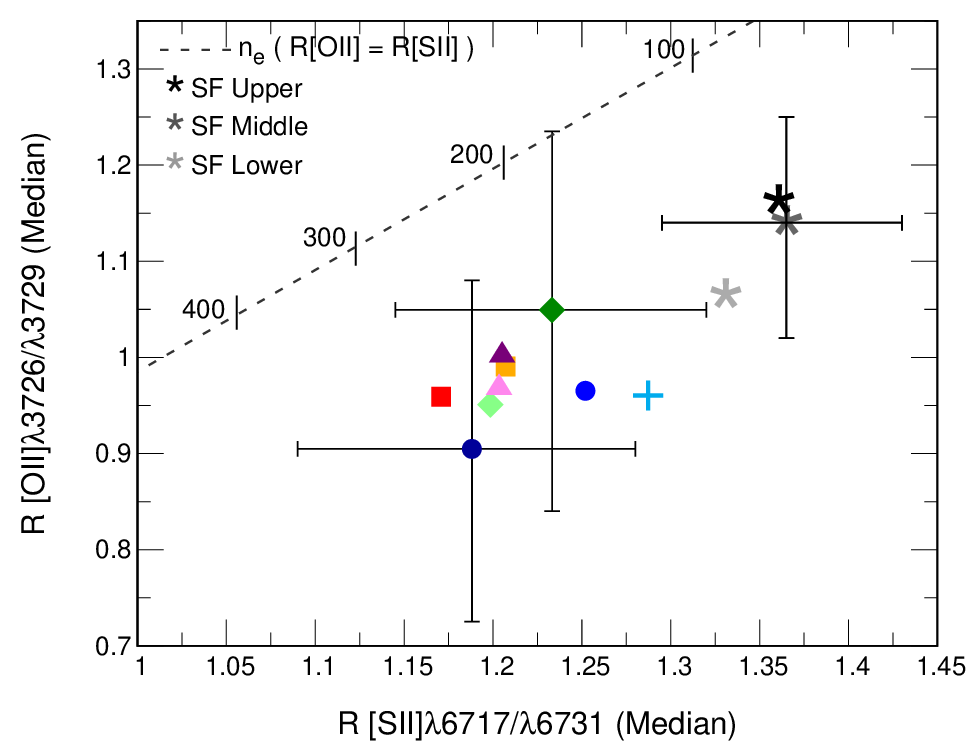}
\caption{ Average electron-density sensitive emission line doublet ratios in classes of galaxy spectra, using same symbols and colors as in previous figures except, with the addition of asterisks for SF galaxies. The dashed line shows the doublet ratios that would result from the same electron density in both the [OII]- and [SII]-emitting regions.
As expected, observed galaxies show higher density gas in their [OII] emission, because it has a higher critical density than the [SII] doublet. Using either observed ratio shows the  trend from lowest electron density in star-forming galaxies, up to highest density in the most AGN-dominated galaxy spectra.  Error bars illustrate the one-sigma scatter of individual galaxy measurements.  Since all of the samples are so large, the resulting uncertainties in the means, i.e. reduced by $\sqrt N$,  are so small that they are comparable to the size of the plotting symbols.}\label{fig:fig10}
\end{figure}

The SDSS spectra resolve two emission 
line doublets, [OII]3726/3729\AA\ and [SII]6716/6731\AA. The observed ratios of these two pairs of collisionally excited lines are sensitive to the electron density of the emitting gas.  The [OII] ratio of 3726/3729 ranges from 1.30 at n$_e$=100 cm$^{-3}$ to 1.14 at n$_e$=400 cm$^{-3}$, while the [SII] ratio of 6716/6731 ranges from 1.31 at
n$_e$=100 cm$^{-3}$ to 1.05 at n$_e$=400 cm$^{-3}$. Because of their different critical densities, the [OII] line emission is more heavily weighted to denser gas than that which emits the [SII] lines. Since the ionized gas in a given galaxy spans a range of electron densities, 
this means that [OII] doublet-determined electron densities are always systematically higher than those determined from the [SII] ratio.  Indeed, we see this effect in each of our galaxy spectroscopic categories, as shown in Figure \ref{fig:fig10}.

Although the [OII]-derived electron densities are higher, they are well correlated with the [SII]-derived densities.  Both doublets show that the ionized gas in SF galaxies is significantly {\it less dense} than that of any Sefyert galaxy class, with the LINERs having intermediate gas densities. Using the [SII] doublet ratio in SDSS spectra, \cite{Zhang2013} also found that an average electron density of 250--300 cm$^{-3}$ in Seyfert galaxies, compared with less than 100 cm$^{-3}$ in SF galaxies.
Furthermore, we find that the more that the Seyfert NLR dominates the spectrum, the higher the average gas density is.  Thus, the Sy 1.0--1.8
galaxies have on average the densest forbidden line-emitting gas (n$_e$[SII] = 250 cm$^{-3}$ and n$_e$[OII] = 1100 cm$^{-3}$.
The Sy2U's have similarly denser gas, whereas the Sy2L's show the effect of some dilution by HII regions. 
Although there is a large intrinsic scatter, we note that for two given Seyfert galaxy spectra, the one which has lower doublet-derived electron densities is more likely to have some emission line contributions from SF regions. 
And even at high redshifts (z$\sim$2), the electron densities in star-forming galaxies remain lower ($n_e$ = 100 - 200 $cm^{-3}$ than in our local AGN \citep{2017ApJ...850..136S,2020ApJ...902..123R}
\footnote{We note that some extreme starburst 
galaxies have very high average electron densities based on their [SII] doublets \citep{2016ApJ...828...67L}.  However even these galaxies, which lie in the SF Upper region, still have lower [OII]-determined electron densities than do the Seyfert-dominated galaxies.}.

Our most important result is that the average Sy1 NLR has higher electron density than the average Sy2, as already first suggested by \cite{1983ApJ...265...92M}.  This observation poses yet another problem for the simple unification picture in which the NLRs of Sy1 and Sy2 are identical.
\footnote{\cite{Vaona2012} also found electron densities from [SII] were about 30\% higher in their Sy1's than in Sy2's. This is shown in their Figure 18, but was not mentioned in their text because it is so subtle, as we also find.} 

\section{Are Weak Seyfert 1's Explained by Low Ionization Parameter, Rather than Star Formation Contamination?
}\label{section7.0}
\begin{figure}[ht!]
\includegraphics[width=0.9\textwidth]{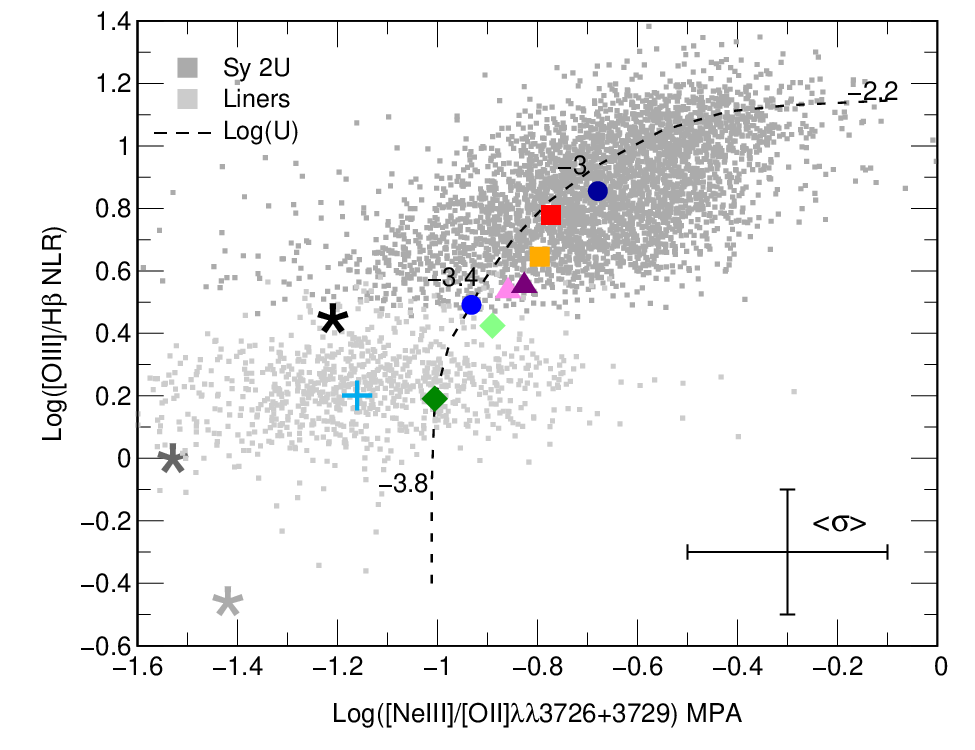}
\caption{Logarithmic line ratios of [O III]/H$\beta$ vs. [Ne III]/[O II (Doublet)]. The dashed line is a sequence of models for the narrow-line region from \cite{2024ApJ...977..187Z}, with the numbers along the line indicating the log of the ionization parameter $U=S_*/(nc)$, where $S_*$ is the flux of ionizing photons and $n$ is the number density of hydrogen atoms. These models are for an ionizing Energy Peak = -1.5, Gas Pressure = 7.4, and standard abundance (12+log(O/H)) = 8.695. The color scheme for identifying different galaxy groups is the same as in previous figures.}
\label{fig:fig11}
\end{figure}

{\it The [NeIII]/[OII] Ionization Diagnostic}. M17 also considered another reddening-independent emission line ratio sensitive to gas ionization, [NeIII]3869/[OII]3726+3729.
Since [NeIII] is not in the Hao data sample, the [NeIII] and [OII]$\lambda\lambda$(3729+3926) are from the MPA-JHU database. The [OIII]$\lambda$5007 and the H$\beta$ narrow line data are from the Hao sample.
As we found for [OIII]/H$\beta$, the AGN narrow lines define a clear sequence with ionization decreasing from 
the BLR-dominated Sy 1.2's and 1.5's to the more SF-dominated Sy 1.8's and 1.9s.
M17 showed that this sequence could be roughly reproduced by a family of AGN photoionization models with varying ionization parameter log U, from \cite{2004ApJS..153...75G}.  These are pure AGN models, with no contribution from SF. 
However, the match with observations was imperfect, since those models tended to over-predict the [NeIII]/[OII] ratio at a given value of [OIII]/H$\beta$.
In Figure \ref{fig:fig11} we have re-done the comparison, using our far larger new dataset, and the latest photoionization models from \cite{2023ApJ...954..175Z,2024ApJ...977..187Z}.
We now see that the new models agree perfectly with the average observed ionization-sensitive line ratios for each AGN class, with log U varying from about -2.6 to -3.6.

The observed fact that the ``weak Seyfert 1's" (Sy1.8 and 1.9) have substantially higher [NeIII]/[OII] ratios than the SF galaxies is one argument that their line emission ratios may be explained by contamination from AGN-photoionized gas with a low ionization parameter. 
The question of how much of the NLR emission in the low-ionization Seyfert spectra is produced by star formation, as opposed to an AGN with lower ionization parameter was also considered by \cite{Richardson2014}. The lower-ionization gas in weak Seyfert galaxies could be produced by NLR gas which is systematically further away from the central nucleus than in strong Seyfert galaxies. Nonetheless, they were unable to rule out an important role for significant emission line mixing from HII regions. Both effects probably play a role in explaining the range of gas ionization levels.
However, there are several stronger arguments that the main differences of the weak Seyfert 1s are {\bf largely} caused by contamination from star formation:

 1. The weak Sefyert 1's are more similar to SF galaxies in all of the line-ratio classification diagrams, and this is quantitatively reproduced with a simple model.
 
 2. The weak Seyfert 1's are more similar to the HII-region dominated galaxies in their gas reddenings and electron densities.
 
 3. In contrast, in most of the observables in 1) and 2), the weak Seyfert 1's are strongly different from LINERs, which are suspected to be dominated by nonstellar photoionization with low U values. For example, their weak [NII], [SII], [OII] and [OI] emission lines cannot be reconciled with a low U value and no SF. Instead, photoionization by young hot stars is favored.
 
 4. There is a natural astrophysical expectation that the {\it relative} importance of SF in spiral galaxies should increase as the power from its AGN decreases.  In contrast, there is no natural prediction of a close correlation between log(U) in the  AGN-photoionized NLR and the relative strength of the BLR.
 
 5. As we will see in the following section, the stellar populations of weak Seyfert 1 spectra also show the strongest signatures of young stars, consistent with the importance of recent SF.

\section{Absorption Line Measurements of Stellar Populations in Seyfert Galaxies}\label{section8.0}

The MPA-JHU catalog of SDSS spectra includes automated measurements for several prominent absorption lines from stellar photospheres. These include the Mgb triplet and the NaD doublet, which are--other things being equal--stronger in cooler
stars, and are therefore strong in the integrated spectra of older stellar populations, whose light is increasingly dominated by red giants. And we include the MPA-JHU measurements of D4000, the drop from 4050--4250\AA\ to 3750--3950\AA. D4000 is a hybrid starlight-and-continuum measure of the combined stellar absorptions of CaII H and K and other blended 
metal absorptions shortward of 4000\AA\ , as well as the overall continuum slope in the violet.  As with Mgb and NaD, D4000 increases as the average age of the stellar population increases. And again, it has a secondary increase if the stellar metallicity increases.

{We plot the strengths of Mg, Na, and D4000 absorption and also the integrated u-g color against the equivalent width of the H$\delta$ absorption line in Figure \ref{fig:fig12}.
Unlike the other absorptions, which are strong in red giants, H$\delta$ is strongest in A stars (e.g., \cite{1983ApJ...265...92M}) . Thus, as the average age of the stellar population decreases, the Mg and Na lines weaken, while the equivalent width of H$\delta$ {\it increases}. In other words, the
stellar population measurements we assembled are sensitive to the star formation history of the galaxy, particularly
SF in the last several hundred million years.  They thus form an {\it age sequence} with average stellar age increasing from upper left to lower right.   There is also a secondary influence on Mg, Na and D4000 absorption, from heavy-element abundance. The higher the metallicity of stars, the relatively stronger are their Mg, Na, and D4000 absorption for a given strength of H$\delta$.
\footnote{The NaD absoprtion suffers from an additional complication, namely that it includes some likely contribution from interstellar gas, as is seen in MaNGA spectroscopy \citep{Machuca2025}. Since we are not able to separate this component, we can only assume that this interstellar NaD is not substantially different in Seyfert and non-Seyfert galaxies. }
\begin{figure*}[ht!]
\includegraphics[width=0.48\textwidth]{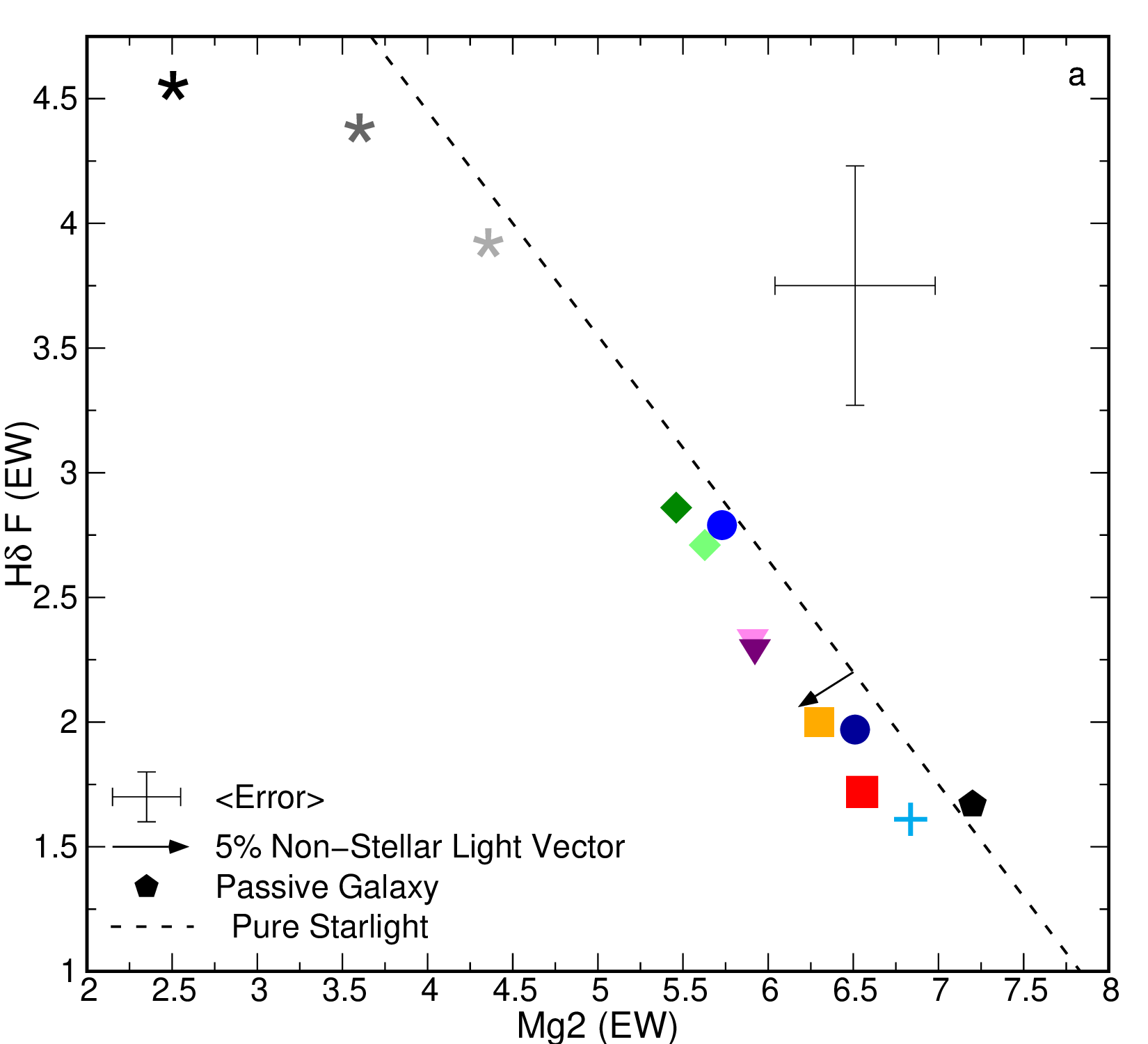}
\includegraphics[width=0.48\textwidth]{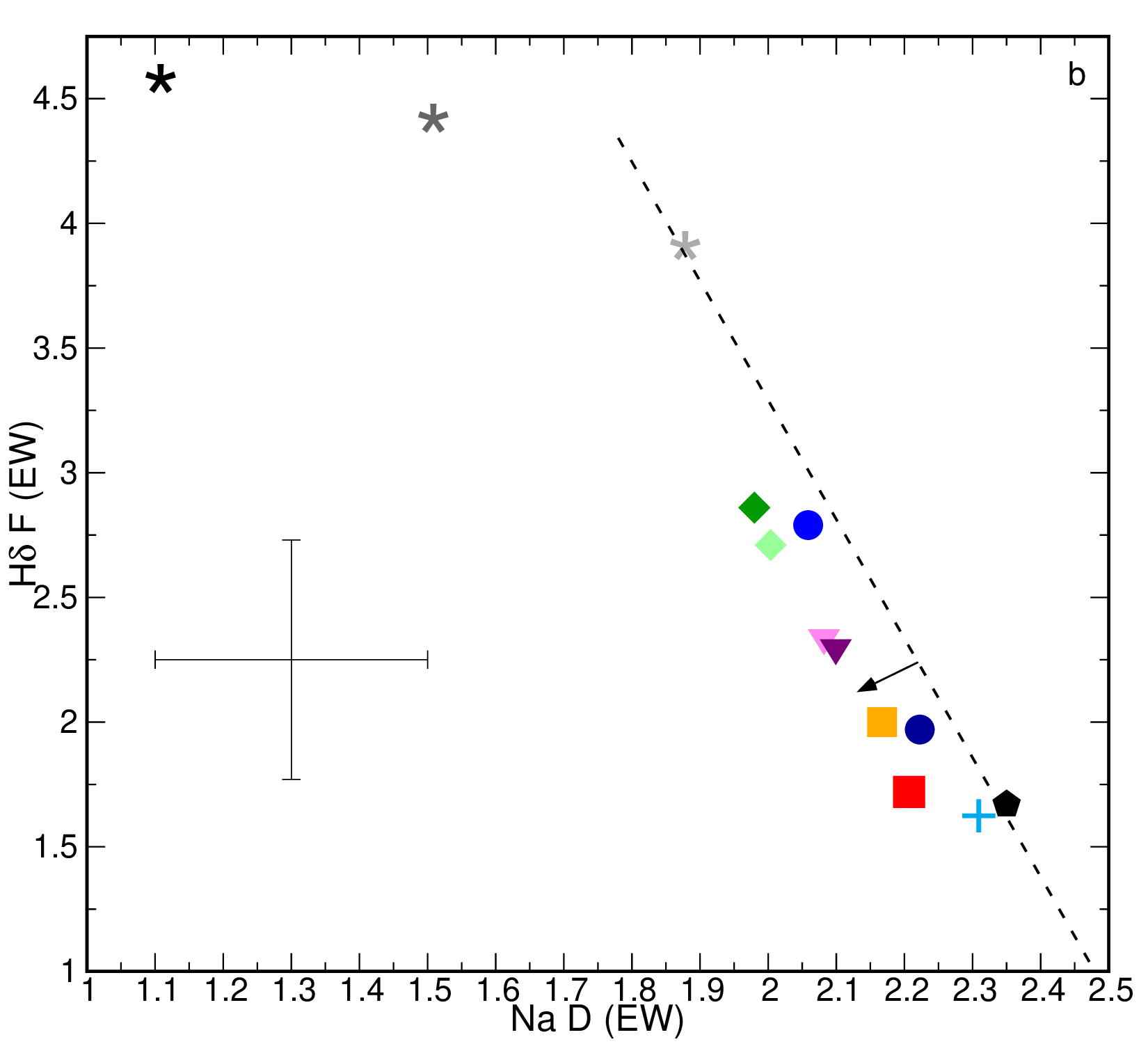}\\
\includegraphics[width=0.48\textwidth]{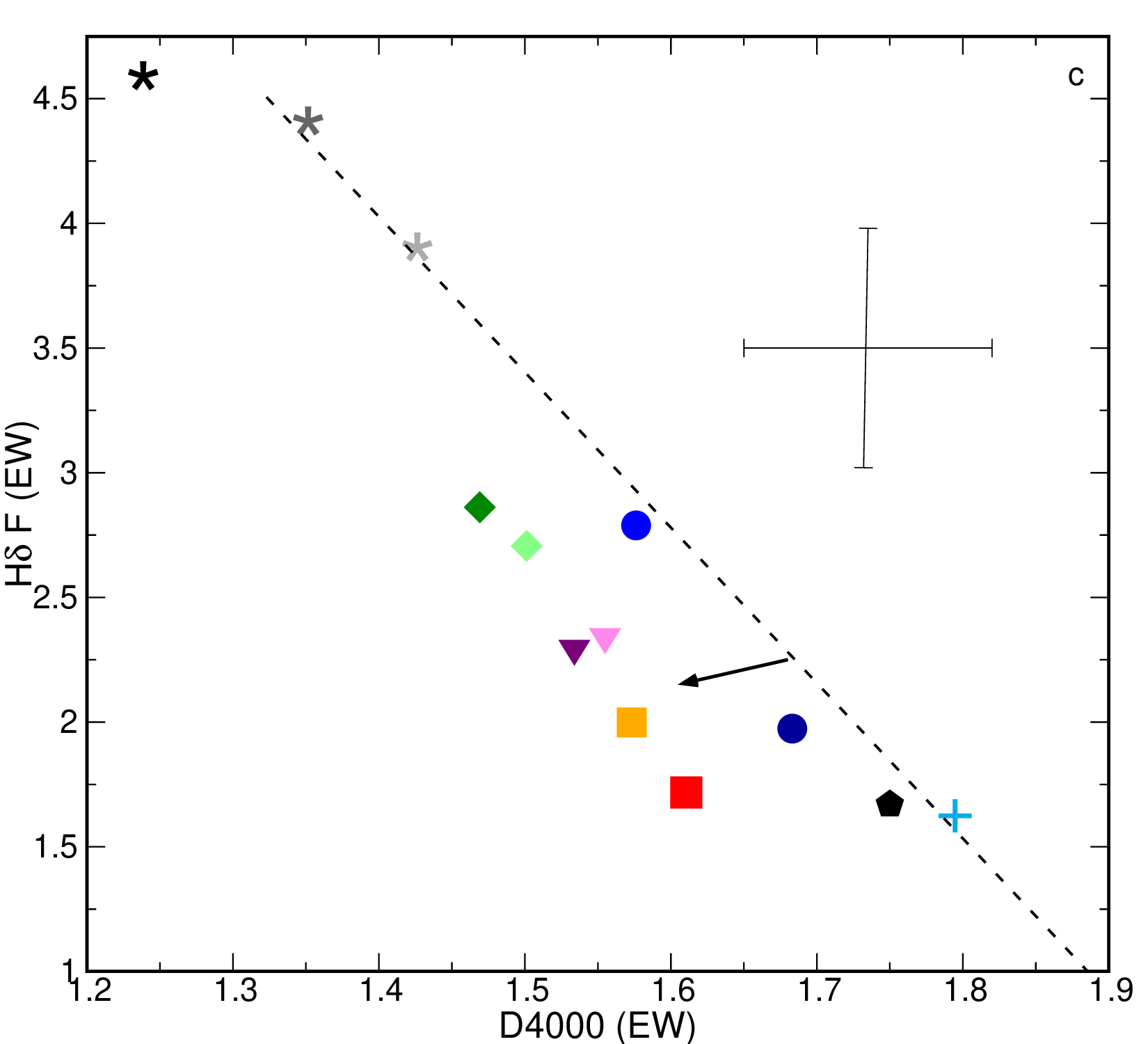}
\includegraphics[width=0.48\textwidth]{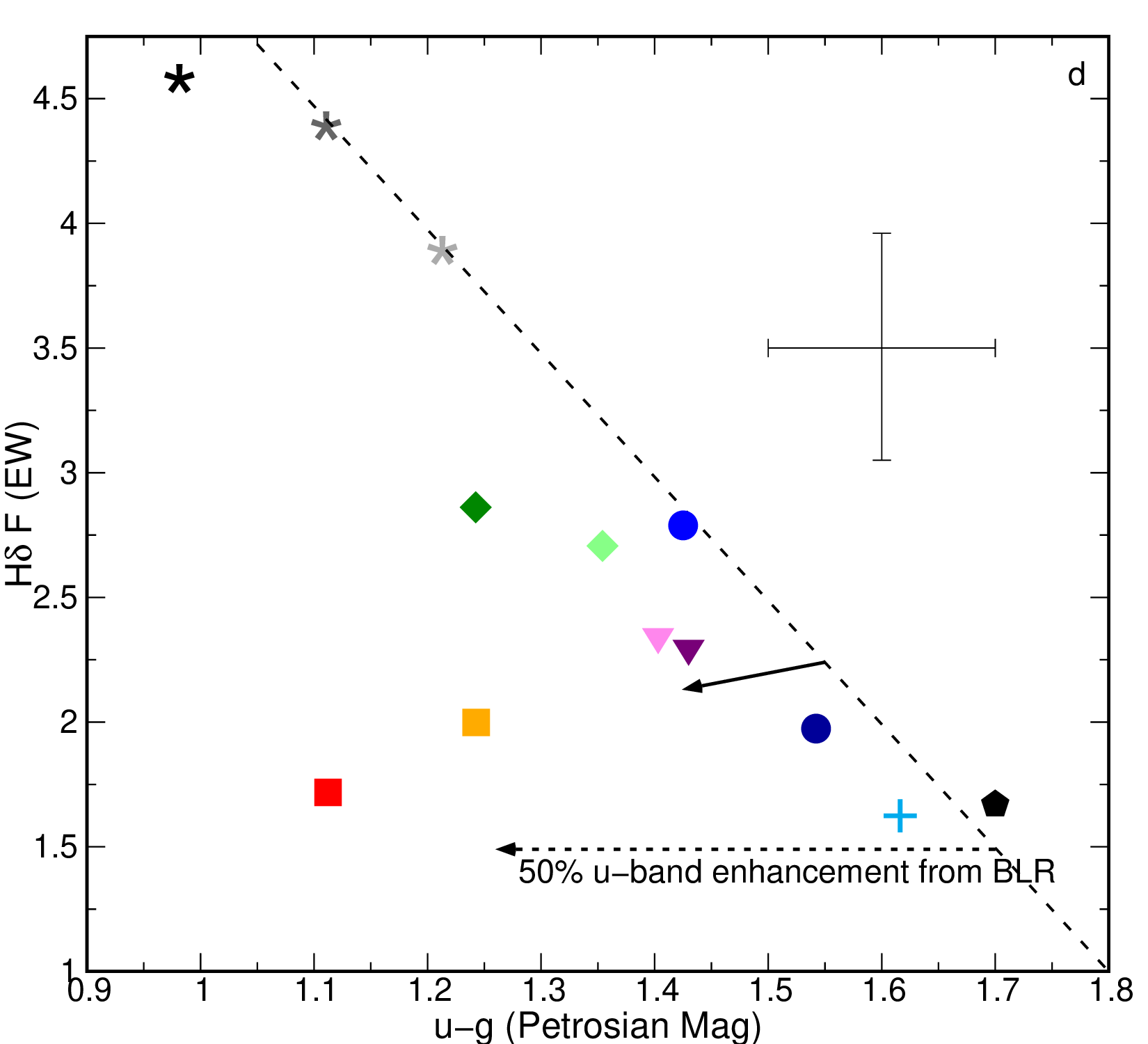}
\caption{Stellar population indicators among the AGN Classes. The horizontal axes show Mg, Na, D4000 absorption line strengths, and integrated u-g color. All of these observables increase with the average age of the stellar population.  In contrast, the vertical axis is always the equivalent width of H$\delta$, which is strongest in young populations with substantial A star contributions.  The points representing means of each galaxy sub-group are shown with the same symbols and colors as in previous figures. The dashed lines show an estimated trajectory for pure stellar populations with age increasing downward to the right, based on the assumption that the continuum spectra of Sy2-Lower and LINER galaxies are purely from starlight.  We assume that LINER galaxies have the oldest stellar populations, with essentially no stars formed in the last Gyr. To confirm this, we have added a point showing average values for 'Passive Galaxies' in each panel. Moving {\it up} this trajectory to the left corresponds to an increasing proportion of younger stars. Shifts to the {\it lower} left of this empirical line (i.e. perpendicularly off of it) allow us to estimate the additional contribution of non-stellar continuum, shown by the black arrows.
Weak Sy1's to have relatively more bluer light due to increased contribution from younger stars.  The effect of adding a 5\% blue contribution from AGN continuum is shown by the solid black arrows. However, the strong Seyfert 1's have much bluer $u-g$ colors which cannot be produced by young stars, since their H$\delta$ absorption is very weak. Instead we attribute this u-bandpass excess to additional contamination from blended Balmer emission lines and continuum produced by the strong BLR.  The estimated average effect of this contamination is shown by the horizontal dashed arrow.}\label{fig:fig12}
\end{figure*}

In Figure \ref{fig:fig12} we show the average values and (individual) standard deviations for our various Seyfert sub-classes. Also plotted with grey asterisks for comparison are the averages for the three groups of star-forming (spiral) galaxies, separated by the ionization level of their gas, with the strongest [OIII]/H$\beta$ ratios corresponding to the lowest metal abundances. We again assume that our normal spiral galaxy categories based on [OIII]/H$\beta$ ratio, SF-Upper, SF-Middle and SF-Lower, correspond to metal-poor, intermediate abundance, and metal-rich galaxies, respectively.

As expected from the considerations above, the x- and y-axis values of galaxies should cluster along a negatively sloped age sequence in Figure \ref{fig:fig12}, with the younger stellar populations found towards the upper left, and the oldest in the lower right.
In all stellar population diagnostic diagrams we find similar trends.
No Seyfert galaxies are as dominated by young stars as much as any of the purely ``star-forming galaxies".
Thus the stellar populations of all Sefyert galaxy types are {\it predominantly old}, with various proportions of young stars mixed in.
 
 Among the various AGN sub-classes, these four diagrams also reveal some further systematic trends.
 The LINERs, which are mostly in early-type galaxies, are found in the reddest, oldest galaxies, very similar to the average of ``passive galaxies" found in the literature \citep{2006MNRAS.373..349R,2002MNRAS.333..517P,2001A&A...366...68P}.
Next, the NLR-dominated Seyfert 2 galaxies (in the ‘upper’ region of the BPT diagrams) have slightly younger stellar populations than LINER galaxies. 
The ‘composite’ Seyfert 2 galaxies (in the ‘lower’ region closer to the star-forming
sequence in the BPT diagrams) are--not surprisingly--in younger galaxies. They fall roughly half way between the oldest galaxies, and the actively star-forming spirals shown by the star symbols.

The Seyfert 1 galaxies follow a very similar systematic trend. 
The strong Seyfert 1's (Sy1.0 and Sy1.2) have old stellar populations, very similar to those of LINERs and pure Seyfert 2's.
The intermediate Seyfert 1's (Sy1.5 and Sy1.6) have significantly younger populations. And the weak Seyfert 1's (Sy1.8 and Sy1.9) have an even larger proportion of younger stars, as is also seen in the composite Sy2's. We have quantitatively confirmed with K-S tests, that each of these 3 groups of Seyfert 1 galaxies has very significantly different average stellar populations. 
The trend of weaker Seyfert 1 galaxies to have younger stellar populations fits well with our finding that their emission lines also show the most significant contributions from HII regions, which are producing ongoing star formation.

{\cite{2003MNRAS.346.1055K} also used the H$\delta$/D4000 diagram to conclude that the stellar populations of many Seyfert galaxies have significant--but not dominant--contributions from relatively younger stars. Another previous study of SDSS spectra of AGN \citep{Marocco2011}  also confirmed our finding that LINERs have the oldest stellar populations (with a mean D4000 of 1.8), while Seyfert 2 galaxies have significantly lower D4000 ($\sim 1.6$), and ``composite" (AGN+SF) galaxies have D4000 values even somewhat lower still. Measuring this diagram for 69 AGN at higher redshifts (z $\sim 1$) in the COSMOS Legacy field, \cite{Mountrichas2022} found the same inverse correlation of H$\delta$ with D4000. In these variables, their AGN have a broad peak around EW(H$\delta$) = 3\AA\ and D4000 = 1.4 (with a long tail up to 1.8).  Although this sample is very small, it confirms our finding that in the past, the stellar populations of AGN at z $\sim$1 included a substantial contribution from young stars. Their proportions may have been somewhat higher than we find in the local universe, as one would expect for any set of galaxies observed at a look-back time of $\sim$ 8 Gyrs.

\section{Small Contamination from Seyfert 1 Nonstellar Light}\label{section9.0}

The Seyfert 1 galaxy spectra show a small but significant displacement down and to the left, ie. perpendicularly away from the stellar population age sequence.  We interpret this as being produced by a small additional continuum contribution from the nonstellar AGN continuum. The black arrow shows the effect of adding a typical AGN continuum contributing 5\% of the total light at 4100\AA. This small featureless AGN continuum partly fills in the stellar absorption lines, making them slightly weaker.  (The AGN contribution may be slightly larger in the strong Seyfert 1's and smaller in the weak Seyfert 1's, but these differences are so minor that they are not statistically significant.)  

The fractional reduction in the stellar absorption line equivalent widths in Figures \ref{fig:fig12} a and b, and \ref{fig:fig12}c is:
\begin{equation}\label{eq:11}
 \Delta (EW) =\frac{p(*)f_{\nu}^* }{p(*)f^{*}_{\nu}+p(AGN)f^{AGN}_{\nu}}\cdot EW(line)
\end{equation}
where the fractional percentages of starlight and AGN continuum are p(*) and p(AGN), which add up to 1.0 at 4100\AA, and $f_{\nu}=C\nu^{\alpha}$, with $\alpha(AGN)=-0.95$ and $\alpha(*)=-1.7$. In each of the diagrams, we find that the Seyfert 1 galaxies typically have 95\% of their continuum produced by an old stellar population, and 5\% by a blue featureless continuum.

The dashed line in each of the Figures \ref{fig:fig12}a--d is a simple straight-line fit for pure starlight being emitted from pure stellar populations in a galaxy, from the top where SF-Lower galaxies are located, down to the LINERs in the lower right of these figures. This sequence is based on the fact that the continuum in star-forming galaxies is dominated by light from hot, luminous young stars, whereas LINER galaxies contain an older stellar population with no star formation in the last Gyr. 
 
In each of these figures SyL and SyU are found somewhat closer to this starlight trajectory than the Sy 1.n.
We attribute this Sy1 offset to a small additional contribution from AGN nonstellar continuum, which is not visible in the Seyfert 2s have their central engine obscured by a dusty tourus, majority of the light from these galaxies is from stars. 

We searched further for evidence of nonstellar AGN light by examining two additional measures of the continuum which are sensitive to violet light, where the AGN contribution is strongest.
We show in Figure \ref{fig:fig12}d the integrated $u-g$ color of the entire galaxy, 
obtained directly from the SDSS SkyServer photometry catalog, averaged over each galaxy class. In Figure \ref{fig:fig12}, both $u-g$ and D4000 are compared against our stellar population age indicator, H$\delta$ absorption.
The same stellar population age sequence is evident.
Again, the $u-g$ color and especially D4000 become smaller in the less AGN-dominated spectra. In other words, the intermediate Sy1's, the Sy2L's, and especially the weak Sy1's show this indication of a larger proportion of blue continuum light from younger stars. Their average stellar populations are significantly younger than those of Passive Galaxies, LINERs, and Sy2U's. 

However, there is an additional effect
evident in the $u-g$ color index: the strong Sy1's (Sy1.2 and especially Sy1.0) shift dramatically to the blue (left), even though their weak H$\delta$ absorption indicates few young stars.
Thus the {\it strong Seyfert 1's have substantial extra flux in the u bandpass, which is not produced by hot stars}.  Instead, we attribute this extra u light to strong BLR emission  contamination. Indeed, 
\cite{1982ApJ...254...22M} first measured the BLR contributions produced by the blended broad-lines of the higher Balmer series above H$\delta$, which then merge into a pseudo-continuum rising smoothly up to 3646\AA, the onset of the Balmer Continuum emission jump.  They found that the Balmer jump can produce as much flux at 3646\AA\ as the underlying continuum.  Based on their finding that the Balmer continuum flux is relatively flat throughout most of the sensitive part of the u filter bandpass (3350--3750\AA), the typical Sy1 studied by \cite{1982ApJ...254...22M} would have its integrated u flux enhanced by a factor of roughly 50\%.  The horizontal vector shows the effect of this BLR contamination on shifting the $u-g$ color of strong Sy1's up to 0.44 mag to the left.  In the intermediate and weak Seyfert 1's, with substantially less broad H$\alpha$ emission, this BLR contamination is too small to be measurable. 
 
Seyfert 1 galaxies are rarely found in high-inclination spiral galaxies, a likely selection effect \citep{1999ApJ...516..660H}.

\section{Selection Effects in Seyfert Galaxy Surveys }\label{section10.0}

AGN which are pre-selected by having quasar-like broadband colors will tend to have relatively stronger nonstellar emission lines (classical NLR and BLR) than AGN turned up in blind spectroscopic surveys of galaxies. The early, highly influential studies of Seyfert 1 galaxies strongly focused on BLR-dominated spectra--our Sy1.0 and Sy1.2, which we call ``strong Seyfert 1's". In other words these are very ``quasar-like" Seyfert galaxy spectra.

Later, large-area mid-infrared surveys such as the 12 Micron Galaxy sample (\cite{1989ApJ...342...83S, 1993ApJS...89....1R}) were sensitive to Seyfert 2 galaxies with emission line spectra dominated by the classical NLR which is powered by the AGN-- almost entirely in the "Upper Seyfert 2" region.

There were, however, hints, that many other AGN would be missed by such focused UV-excess and IR continuum searches.
Radio observations suggested that a small minority of AGN were ``radio-loud", and these sometimes lacked other properties of AGN-dominated Seyferts \citep{1996ApJ...473..130R}. Similarly, surveys of hard-X-ray sources included many previously overlooked AGN \citep{1996ApJ...471..190R, 2022ApJ...936..162S}.

The effects of selecting AGN independent of their nonstellar UV and IR continuum emission became evident when large scale spectroscopic surveys were made without such pre-selection. 
For example, the Seyfert 2 galaxies in the CfA sample include larger contributions from star formation, i.e., their spectra are less AGN-dominated \citep{1987ApJ...321..233E}.
These new AGN samples included more significantly reddened nuclei, with little or no UV-excess, and large continuum contributions from starlight. They tend to be what are now called ``Composite" AGN (M17).

 This trend to find more of the numerous Sy1s with weaker BLRs, which we call Sy1.6, 1.8 and 1.9, continued with the SDSS spectroscopy. These changing selection methodologies explain why
 pioneering studies of Sy2's refer largely to what we call ``Seyfert 2 Upper" AGN, while Sloan now selects large numbers of ``Seyfert 2 Lower" AGN.
The new Sloan Seyfert 2's are less AGN-dominated, and, we suspect based on incomplete data, very unlikely to harbor hidden broad line regions that could be uncovered by spectro-polarimetry (\cite{2008ApJ...676..836T}).

\subsection{``Strong Unification" of Seyfert 1 and 2 Galaxies} \label{section10.1}
In the most popular version of Seyfert ``unification", all of the {\it apparent} differences between Sy1 and Sy2 AGN can be explained by our viewing angle with respect to a hypothetical dusty ``torus" \citep{1998ApJS..117...25M, 2014ApJ...788...45T}. Since this torus is larger than the BLR but smaller than the NLR, strong unification predicts no systematic differences between
the NLR's of Sy1 and Sy2 galaxies.  Indeed the small AGN samples in M17 did not show strong differences. 
And further, if we restricted our consideration to only those Seyfert 1 and Seyfert 2 galaxy spectra which are strongly dominated by the AGN, they do occupy the same locations in the BPT line-ratio classification diagrams.

Even when we restrict consideration to the minority of AGN-dominated Seyfert galaxy spectra, our current much larger samples do actually show that Sy2 nuclei have--on average--significantly higher NLR reddenings than do Sy1's, and especially Sy1's with strongly dominant BLR components. This difference was also identified by \cite{1998ApJS..117...25M}, who claimed the Sy2's have on average additional extinction from dust which is more spatially extended than the NLR,
and is sometimes observed directly by HST imaging as nuclear dust lanes. These are less often present in Sy1 galaxies
\citep{2004ApJ...616..707H}.

When we extend our consideration to all Sy 1 and 2 galaxies, the 
simple unification picture breaks down further, because of an {\it important new parameter}--the contribution of star formation to the observed NLR fluxes. We have shown that these SF contributions can be large or small.  Unless the SF contributions of the host galaxies could somehow be matched, the strong unification prediction of identical NLR emission in Sy1 and Sy2 samples will not in general be observed.

\subsection{Are All Broad Line Regions the Same?}
Using extensive spectroscopic data, we have made detailed comparisons of the Narrow Line Regions in all types of Seyfert galaxies. Throughout this analysis, we have implicitly {\ it assumed} that the Broad Line Regions producing the H$\alpha$ emission with velocities of thousands of km/sec are the same in all Sy1s. We have measured one observable which can test this assumption--the velocity widths of the various sub-classes of Seyfert 1 galaxies.  We find that the median FWHM of broad H$\alpha$ components in our weak and intermediate Seyfert 1 galaxies are all about 2400 km/sec. However, the BLR velocities in strong Seyfert 1s are somewhat larger.  The median FWHM of broad H$\alpha$ in our Sy1.2's is 3400 km/sec, and in Sy1.0's it is 4400 km/sec. A slightly smaller increase in average FWHMs was found by \cite{Oh2015} in their measurements of Sy1's with weak and strong broad H$\alpha$ components.  It is possible that part of this difference is a selection effect, in which broader H$\alpha$ wings are more detectable above the continuum when the relative amplitude of the broad Gaussian is larger.  However, if some of the differences are intrinsically real, it suggests that more luminous BLR's tend to have somewhat higher velocity dispersions. One speculation could be that more of their ionized hydrogen is relatively closer to relatively more massive central black holes. Measuring--and possibly explaining--this apparent trend will require analyzing spectra that go well beyond the scope of our current study. In any case, these possible BLR velocity differences do not impact the various NLR properties and differences that we described above.

\section{Conclusions}\label{section11.0}

We have measured the narrow optical emission lines in the spectra of a very large sample of Seyfert 1 galaxies, along with comparison samples of Seyfert 2 galaxies, LINERs and star-formation dominated galaxies.

By analyzing several NLR classification diagrams, we find that the narrow line emission in Sy 1 nuclei with the weaker relative broad line components--what we call Sy 1.5, 1.6, 1.8 and 1.9's--is partly produced by HII regions, instead of the classical Sy 2 AGN-photoionized NLR.  This hypothesis is further supported by similar systematic differences we find in gas reddening and electron density between the galaxy categories: the strong Seyfert 1's are most like pure AGN, while the weak Seyfert 1's are more like the star-forming galaxies. We present a simple AGN+SF mixing model which reproduces these observations.

Our analysis of the stellar absorption lines in these Seyfert spectra confirms this trend. The AGN-dominated spectra (including Sy 1.0s and 1.2s) have predominantly old stars, similar to those present in "pure" Sy2's and LINERs. In contrast, in the Sy 1.5--1.9 sub-groups the stellar populations also have a quite substantial additional contribution from young stars.
In the most broad-line-dominated Seyfert 1 spectra, we also find a strong component of violet continuum which we attribute to the blended Balmer lines and continuum from the BLR.
 
Although our Seyfert samples were based on  line ratio diagrams to exclude LINERs, our large sample does reveal a minor contribution to the Seyfert emission lines from a LINER component, although much smaller than the contribution from HII regions. The LINER presence is strongest in the intermediate Seyfert 1 galaxies, our types 1.5 and 1.6; nearly 15\% of these turn out to be LINER-dominated. 

Simplistic tests of Type 1 and Type 2 AGN "Unification" schemes, which assume that all NLR emission in all AGN is the same, will not in general be valid. More realistic models must take into account the widely varying amount of spectroscopic contamination produced by young stars and their associated HII regions. 

\section{Acknowledgement}\label{Acknowledgement}

We gratefully acknowledge Chenxu Liu for help with the multiple-Gaussian fitting of the emission line blends, used in Figure \ref{fig:fig1}. 
We thank Peixin Zhu, for sharing with us the numerical results of her recent AGN photo-ionization model calculations. 
We thank the anonymous Referee for many insightful comments and suggestions which significantly improved this paper.
Funding for the Sloan Digital Sky Survey V has been provided by the Alfred P. Sloan Foundation, the Heising-Simons Foundation, the National Science Foundation, and the Participating Institutions. SDSS acknowledges support and resources from the Center for High-Performance Computing at the University of Utah. SDSS telescopes are located at Apache Point Observatory, funded by the Astrophysical Research Consortium and operated by New Mexico State University, and at Las Campanas Observatory, operated by the Carnegie Institution for Science. The SDSS website is \url{www.sdss.org}.

SDSS is managed by the Astrophysical Research Consortium for the Participating Institutions of the SDSS Collaboration, including the Carnegie Institution for Science, Chilean National Time Allocation Committee (CNTAC) ratified researchers, Caltech, the Gotham Participation Group, Harvard University, Heidelberg University, The Flatiron Institute, The Johns Hopkins University, L'Ecole polytechnique f\'{e}d\'{e}rale de Lausanne (EPFL), Leibniz-Institut f\"{u}r Astrophysik Potsdam (AIP), Max-Planck-Institut f\"{u}r Astronomie (MPIA Heidelberg), Max-Planck-Institut f\"{u}r Extraterrestrische Physik (MPE), Nanjing University, National Astronomical Observatories of China (NAOC), New Mexico State University, The Ohio State University, Pennsylvania State University, Smithsonian Astrophysical Observatory, Space Telescope Science Institute (STScI), the Stellar Astrophysics Participation Group, Universidad Nacional Aut\'{o}noma de M\'{e}xico, University of Arizona, University of Colorado Boulder, University of Illinois at Urbana-Champaign, University of Toronto, University of Utah, University of Virginia, Yale University, and Yunnan University

\bibliographystyle{aasjournalv7}

\end{document}